\documentclass[useAMS,usenatbib]{mn2e}

\usepackage{graphicx}       		

\def\hb{H$\beta$}
\def\te{T$_e$}
\def\ne{N$_e$}
\def\tenii{T$_e$[N~{\sc ii}]}
\def\teoiii{T$_e$[O~{\sc iii}]}

\def\nesii{N$_e$[S~{\sc ii}]}

\title[The spectroscopic mapping of NGC~40 using {\sc 2d\_neb}]{Physico-chemical spectroscopic mapping of the 
planetary nebula NGC~40 and the {\sc 2d\_neb}, a new 2D algorithm to study ionised nebulae}
\author[M. L. Leal-Ferreira, D. R. Gon\c calves, H. Monteiro and  J. W. Richards]
{M. L. Leal-Ferreira$^{1}$\thanks{Currently at Argelander-Institut f\"ur Astronomie, University of Bonn, Auf dem H\"ugel 71, D-53121 Bonn, Germany. E-mail:
ferreira@astro.uni-bonn.de (MLLF)}, 
D. R. Gon\c calves$^{1}$, H. Monteiro$^{2,3}$ and J. W. Richards$^{4,5}$ \\
$^{1}$UFRJ - Observat\'orio do Valongo, Ladeira do Pedro Ant\^onio 43, 20080-090, Rio de Janeiro - RJ, Brazil\\
$^{2}$N\'ucleo de Astrof\'isica Te\'orica-CETEC-UNICSUL, Rua Galv\~ao Bueno, 868, 01506-000, S\~ao Paulo - SP, Brazil\\
$^{3}$Universidade Federal de Itajub\'a, Av. BPS, 1303, Itajub\'a - MG, Brazil\\
$^{4}$Astronomy Department, 601 Campbell Hall, University of California Berkeley, Berkeley, CA 94720, USA\\
$^{5}$Statistics Department, 367 Evans Hall, University of California Berkeley, Berkeley, CA 94720, USA}
\begin{document}

\date{Accepted...; in original form...}

\pagerange{\pageref{firstpage}--\pageref{lastpage}} \pubyear{2010}

\maketitle

\label{firstpage}

\begin{abstract}
In this paper we present an analysis of the physical and chemical conditions of the planetary nebula 
NGC~40 through spatially-resolved spectroscopic maps. We also introduce a new algorithm --{\sc 2d\_neb}--
based on the well-established {\sc iraf} \emph{nebular} package, which was developed to enable the use of 
the spectroscopic maps to easily estimate the astrophysical quantities of ionised nebulae. 
The {\sc 2d\_neb} was benchmarked, and we clearly show that it works properly, since it compares nicely 
with the {\sc iraf} \emph{nebular} software. 

Using this software, we derive the maps of several physical parameters of NGC~40. From these maps, we conclude 
that \tenii ~shows only a slight temperature variation from region to region, with its values constrained 
between $\sim$8,000~K and $\sim$9,500~K. Electron densities, on the other hand, have a much more 
prominent spatial variation, as \nesii\ values vary from $\sim$1,000~cm$^{-3}$ to $\sim$3,000~cm$^{-3}$. 
Maps of the chemical 
abundances also show significant variations.  From the big picture of our work, we strongly suggest 
that analysis with spatial resolution be mandatory for more complete study of the physical and 
chemical properties of planetary nebulae. 
\end{abstract}


\section{Introduction}
 
  Stars of low and intermediate mass (0.8 - 8~M${_\odot}$) represent
  95\% of the evolved stars in the Galaxy and, because of the initial
  mass function, can be important sources of enrichment of He, N and C
  in the Interstellar Medium (ISM) (e.g., \citealt{yin10}). Although 
  supernovae are expected to be the main source of enrichment for most 
  species \citep{chiappini03}, \citet{hoek97} and \citet{marigo01} 
  predicted theoretically the possibility that planetary nebulae (PNe)
  progenitor stars also produce O and Ne and dredge these elements up to
  the stellar surface, resulting in possible enrichment of the ISM. In
  any case, PNe and the study of their abundances are clearly
  important tools to study stellar evolution as well as --by considering
  elements other than those just discussed-- revealing the
  imprint of ISM abundances when their progenitor stars were formed,
  in the Galactic as well as extragalactic context.

The outer layers of the intermediate mass stars, which were enriched
throughout their evolution, are ejected during the Asymptotic Giant
Branch (AGB) and post-AGB phases, and subsequently ionised by the
remnant hot nucleus, referred to as the central star (CS). Analysis of
the ionization lines formed in the ejected envelopes of these evolved
stars is one of the most important ways in which the chemical
and physical characteristics of the gas are studied. However, the
great majority of work done in this field utilizes single slit
spectroscopy to measure the ionization lines and thus is limited to
obtaining average values across the observed region of the slit (or
across multiple regions, as it is sometimes done). Utilizing single
slits is extremely limiting in the analysis of these objects,
especially considering the high complexity of PNe morphology.  In
single-slit analysis, the possible spatial variations of PNe are
usually ignored.

However, spatial variations in the chemical and physical conditions of PNe 
have been detected in previous works using long slits. \citet{guerrero96},
for example, using long-slit spectra, found that the N/O ratio varies
from 0.4 to 4.2 from one region to another in the planetary nebula
K~4-55. Moreover, also using long-slit spectra, \citet{balick94} found
N/H abundance gradients in a few PNe. In particular they claimed to
have found overabundance factors of 2 for NGC~6826 and NGC~6543 and 5
for NGC~7009, respectively. On the other hand, another empirical study of 
NGC~7009 \citep{goncalves03} lowered the overabundance factor to at most 2, 
and a detailed 3D photoionization modelling of the nebula eliminated
completely the need of any overabundance to explain the observed
long-slit as well the narrow-band images of the nebula
\citep{goncalves06}. By investigating 13 bipolar PNe using long-slit
spectrophotometry, \citet{perinotto98} concluded that (within the
errors) He, O, and N abundances are constant throughout the nebulae,
and that the Ne, Ar, and S abundances are also constant, but their face
values have systematic increases toward the outer regions of the
nebulae. The latter was attributed to inaccuracies in the ionization
correction factors.

In recent years a new tool for obtaining spectroscopic data has been
developed. The Integral Field Unit (IFU) is a tool
that utilizes a square array of lenses remapped to a slit and then
imaged through a spectroscope to obtain spatially resolved data
\citep{ren02}. \citet{tsamis08} were the first to use this tool to
investigate the spatial variations of PNe parameters. They verified
that there is variation on the discrepancy of the O$^{++}$ abundances
obtained from the collisionally excited lines (CEL's) and optical
recombination lines (ORL's) on the studied PNe. They established
correlations between the ionization state of the gas and the
discrepancies of O$^{++}$ for 2 PNe and between the discrepancy of the
O$^{++}$ and the abundances of C$^{++}$ and O$^{++}$ with the
electron temperature for all 3 studied PNe.

Similar kinds of data can be produced by a spectroscopic mapping
technique developed by \cite{b10}, in which multiple parallel
long-slit spectra of a nebula are interpolated to create emission-line
maps. Both types of data, those obtained from IFU and those found by the 
spectroscopic mapping technique, enable better spatially-resolved analysis of
ionised nebulae than those based on single long-slit
spectra. Spatially-resolved data are also critical in constraining
photoionization models which are now fully three-dimensional and can
produce a great number of observable quantities, including
projected images for each emission line. Photoionization models that
utilize these constraints were obtained by \cite{b10},
\cite{b14}, \cite{M06} for {NGC~6369}, {Menzel 1} and {NGC~6781} as
well as by \cite{MG09}.

To further our understanding of the spatial variations of the physical and
chemical properties of PNe, we present in this work results of the
study of NGC~40 using spectroscopic mapping. Following the
morphological classes of \citet{balick87}, from their H$\alpha$,
[N~{\sc ii}] and [O~{\sc iii}] images, the PN NGC~40 has an elliptical
shape, which can also be seen in the H$\alpha$ narrow band image
obtained in our observing run (Figure~1). Two distinct structures are
found within the main elliptical region: a bright inner rim and an 
outer shell.  A filament-like structure can also be identified beyond
the outer shell, on the northeast region. The [O~{\sc iii}] emission,
originating in a hotter, low-density medium, is located between the hot
bubble and the H$\alpha$ (or [N~{\sc ii}]) filaments, and is better
represented by a ring like structure \citep{balick87}.

In X-rays NGC~40 also shows a ring-like emission region, and the
correspondence of this region with the inner rim indicates that the
bright envelope, seen in the optical image, is caused by the
shocked CS fast wind. The absence of X-ray emission in
the region of an apparent aperture in the inner rim suggests 
that this wind is quite spherical \citep{montez05}. Based on UV data, 
\citet{b4} suggest that the C~IV 1549~\AA ~emission from the nebular 
envelope is too strong to have been produced by normal thermal processes, 
and that it is probable that this emission is a consequence of processes 
related to the central star wind.

We note that there is a discrepancy in temperature determinations for
the CS of NGC~40, a Wolf-Rayet (WC8), in the literature. \citet{b20}
calls our attention to the peculiarity of the fact that NGC~40 is a
low-excitation PN but seems to be excited by a 90,000~K central
star. However, a temperature of about 40,000~K, which would be
consistent with a low excitation PN, is usually adopted for the NGC~40
central star \citep{stankevich,b17}.  \citet{b21} argue that a
``carbon curtain'', located around the CS, could be the mechanism
responsible for hiding the CS true effective temperature.

A number of works have dealt with the physical parameters and
chemistry of NGC~40 in the past decades (\citealt{b17}; \citealt{b8}; 
\citealt{b9}). These authors investigated different regions of the nebula 
utilizing inconsistent slit configurations or apertures, and thus 
found results that differ greatly from one paper to another. Even 
though these works provide a reliable overall picture of the object, they fail 
to give more detailed information that could potentially lead to important 
clues as to shaping mechanisms and enrichment patterns, among others.

One important practical limitation in dealing with spatially resolved
data is the fact that all currently-available tools used to estimate the
physical and chemical properties of PNe have been limited to single
numerical values consistent with integrated long-slit fluxes. An 
alternative to the above could be to implement scripts written for {\sc iraf}, 
which would derive the physical quantities in a pixel per pixel, or region 
per region fashion, and then to arrange all the results together. However, 
these kinds of analyses have rarely been reported.

In this work we present a tool, based on the well established
{\sc iraf} {\it nebular} package (stsdas.analysis; De Robertis, Dufour
\& Hunt 1987), which is capable of working with two-dimensional data,
such as IFU and spectroscopic
maps. 
This tool, the {\sc 2d\_neb} package, has been tested and shown to be
consistent with results obtained with {\sc {\sc iraf}} {\it nebular}
procedures (see Sec. 3.4). We also utilize the {\sc 2d\_neb} package
to study the PN NGC~40 and present the first spatially resolved results of the
spectroscopic mapping of this nebula, for which we have obtained the
extinction coefficient, electron density and temperature, and ionic
abundances of several elements.

This paper is structured as follows. In Section~2 the acquisition and
reduction of the data are discussed. Section~3 presents the method
of data treatment, including a description of the spectroscopic mapping 
technique and explanation of those algorithms included in the {\sc 2d\_neb} 
package that were developed to analyse spectroscopic 
maps. Some benchmarks of 
this new package are also given within Section~3. In Section~4 the empirical 
results for one of the long-slits observed in this work are compared with those 
obtained from the spectroscopic maps by means of the {\sc 2d\_neb} package. 
Section~5 is dedicated to our spatially resolved results for NGC~40, 
while the summary and conclusions are in Section~6.

\section[]{Data Acquisition and Reduction}

The observational data used in this work includes both an H$\alpha$
image and optical long-slit spectra. The observations were obtained on
the night of 2005 October 28$^{th}$, at the 2.56~m Nordic Optical
Telescope (NOT), located at the Observatorio del Roque de los
Muchachos (European Northern Observatory, La Palma, Spain). For both the
image and spectra, the Andalucia Faint Object Spectrograph and Camera
(ALFOSC), which has a pixel scale of 0.189\arcsec~pixel$^-$$^1$, was
used.

The H$\alpha$ image has an exposure time of 30~s and the seeing during
the exposure was 0.89\arcsec. In this paper this image is used only
for comparison with the H$\alpha$ spectroscopic map that is
constructed using the spectra, as described in detail below. The
reduction of the H$\alpha$ image included only bias subtraction, in 
addition to flat-field (dome and sky) correction. No photometric
calibration of this image was performed.  It is worth noting,
however, that the resolution of the H$\alpha$ image is better than the
resolution achieved with the maps.

\begin{figure}
\begin{center}
\includegraphics[width = 0.48 \textwidth]{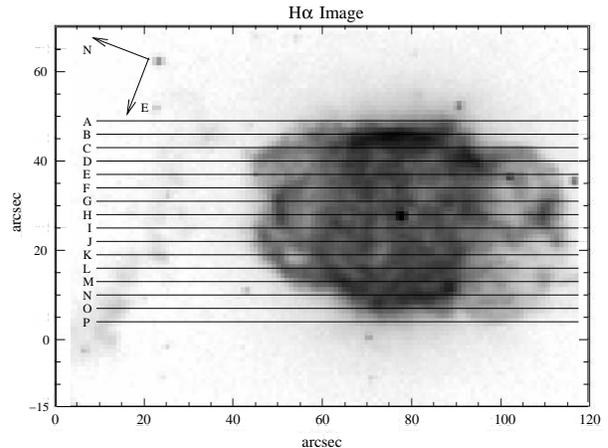}\\
\caption[]{NGC~40 image, observed with an H$\alpha$ filter. The parallels 
slit positions are represented by lines and labeled by letters.
}
\label{imhalpha}
\end{center}
\end{figure}

For spectroscopy, the long-slit width used was 1.3\arcsec. The
ALFOSC was used with the grism number 7, which has 600 rules
mm$^-$$^1$ and a spectral coverage from 3850~\AA~to 6850~\AA. The
reciprocal dispersion of the binned pixel is
3.0~\AA~pixel$^-$$^1$. The slit was located on 16 different parallel
positions across the nebula.  The distance between them was fixed and
set at 3\arcsec. For 15 of those positions the exposure times were 
3 $\times$ 300~s. For slit B only one 300~s exposure was
taken. Figure~\ref{imhalpha} shows our NOT H$\alpha$ image, and also
the relative slit positions across the nebula. Note, in addition, that
the slits do not cover the entire object.  Slit lengths in
Figure~\ref{imhalpha} are reduced for visualization purposes with 
their respective labels used throughout the text.

The data reduction was performed using the standard procedures for
long-slit spectroscopy of the Image Reduction Analysis Facility ({\sc
  {\sc iraf}}).  Bias frames, flat-field, helium-neon wavelength
calibrations and standard star (G191B2B) exposures were obtained
during the observation runs, and used in the reduction/calibration
procedure. Three spectra per slit position were taken to improve the
signal-to-noise ratio (S/N) and eliminate cosmic rays. The only
non-standard procedure used was binning on the dispersion axis. This
was necessary to improve the wavelength ($\lambda$) calibration due to
the fact that the helium-neon lamp was resolved in the arc frames.

\begin{figure}
\begin{center}
\includegraphics[width = 0.47 \textwidth]{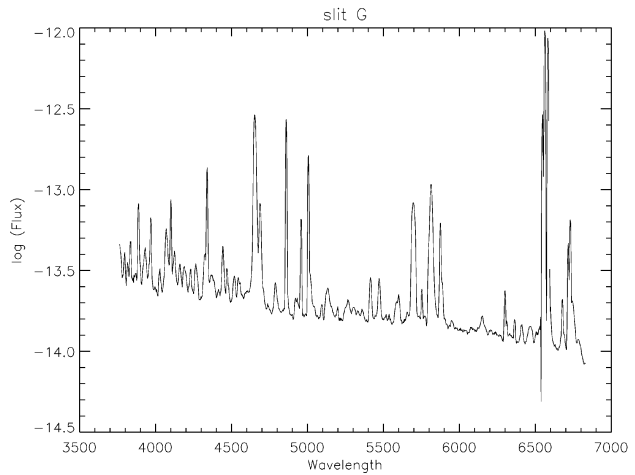}\\
  \caption[]{Spectrum of the NGC~40, integrated over the whole slit G.}
\label{slitspectra}
\end{center}
\end{figure}

The final calibrated spectrum for slit G, which passes through the
central star, is shown in Figure~\ref{slitspectra}. Fluxes here are
representative of the emission integrated along the slit, which we will
later call ``WN'' (whole nebula) when comparing long-slit measurements and 
results obtained from the spectroscopic maps, in Sections 3.2 and 4.

\section{2D Observations and Tools for Gaseous Nebulae Analysis}
\subsection{Spectroscopic Mapping Construction}

\begin{figure}
\begin{center}
\includegraphics[width = 0.48 \textwidth]{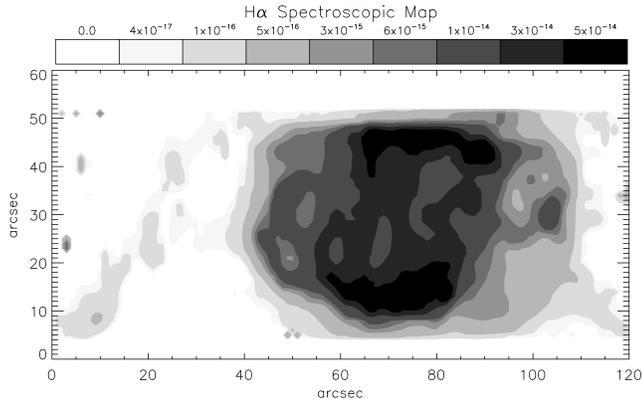}\\
\caption[]{H$\alpha$ spectroscopic map of NGC~40, obtained with the 
spectroscopic mapping reconstruction technique. Fluxes are given in 
erg~cm$^{-2}$~s$^{-1}$. }
\label{hacontour}
\end{center}
\end{figure}

Emission-line mapping is a technique in which the spatial profiles
from a set of parallel slits, for a given emission-line, are combined
to create an emission-line map of the nebula \citep{b10}. We applied
this technique using the 16 parallel long-slit spectra described
above, to obtain the emission-line maps of NGC~40.  These maps correspond to a
field of view of 120\arcsec\ $\times$ 47\arcsec\ (see
Figure~\ref{hacontour}).

This method has been previously applied to  Hubble~5, NGC~6369, Mz~1
and NGC~2022 (\citealt{b13}; \citealt{b10}; \citealt{b14} and \citealt{b15},
respectively), providing informative quantities such as the total emitted 
flux of each object and variations of density and temperature across the 
nebulae, among others. For example, \citet{b14} found indication of the
presence of a dense ring and of a bipolar structure of lower density
for the planetary nebula Mz~1. Based on that map, they proposed a
tridimensional hourglass model to represent the gas distribution of
the nebula.

In the present paper, the above method was applied to generate 31
spectroscopic maps of the PN NGC~40, each corresponding to a 
different emission line. The generated maps were
corrected for the effects of field rotation due to the Atl-Azimuth 
nature of the telescope as well as for differential atmospheric refraction,
using the tables of \cite{F82}. To avoid the presence of pixels with
spurious information, all created maps have been submitted to a signal
to noise ratio (SNR) cut. These cuts were done by applying a mask in
which all the pixels with a SNR lower than a certain value were
artificially substituted by 0 (zero). For the brighter emission-lines,
the adopted value was SNR equal to 7. On the other hand, this limit
did not work for the weaker emission-lines, as almost no signal would
survive to the noise-mask cleaning. To handle this problem, noise-masks 
based on different values were tried. Table~\ref{mapintens} presents 
the list of all the 31 emission-line maps that were constructed, 
and the correspondent SNR cut-offs ($SN_{cut}$) of these maps. The  
third, fourth and fifth columns of this table show, respectively, the 
integrated spectroscopic map fluxes ($F_{\lambda}$) --corresponding to the 
emission of the entire observed nebula--, the extinction-corrected intensities
($I_{\lambda}$, as will be discussed in Section~5.1), and the number of the 
figure in which the map is presented in this paper (Fig.).

Figure~\ref{hacontour} shows the H$\alpha$ spectroscopic map. As we noted 
in Figure~\ref{imhalpha}, the fact that slits do not cover the entire 
nebula is clearly seen here: the bright edges determine the limits of our
observations and correspond to the slits A and P. It is evident from 
Figure~\ref{hacontour} that the H$\alpha$ structures of the image in 
Figure~\ref{imhalpha} are well-reproduced by the spectroscopic map. The 
latter apply not only to the bright elliptical nebular components --the 
inner rim and outer shells--, but also to the much fainter structures of 
smaller scales, as well as to the extended filaments.

\begin{table}
  \centering
  \begin{minipage}{76mm}
    \caption{List of the 31 spectroscopic maps with the corresponding
      SNR cut-offs.  Observed fluxes and intensities
      (extinction-corrected) are normalized to H$\beta$ = 100.}
    \begin{tabular}{@{}lllll@{}}
      \hline
      Line Identification & $SN_{cut}$ & $F_{\lambda}$ & $I_{\lambda}$ & Fig.\\
      \hline
      H10~3797 & 7.0 & 3.15 & 3.98 &  -- \\
      H9~3835 & 7.0 & 6.48 & 7.27 &  -- \\
      H8$+$He~{\sc I}~3888 & 7.0 & 15.40 & 17.24 &  -- \\
      H$\epsilon$$+$$[$Ne~{\sc iii}$]$$+$He~I~3968 & 7.0 & 11.38 & 13.79 &  -- \\
      $[$S~{\sc ii}$]$$+$$[$S~{\sc ii}$]$~4069+76 & 4.0 & 0.64 & 0.73 &  -- \\
      H$\delta$~4101 & 7.0 & 18.97 & 22.43 &  -- \\
      H$\gamma$~4340 & 7.0 & 38.81 & 43.85 & --  \\
      He~{\sc i}~4388 & 2.5 & 0.84 & 0.05 &  -- \\
      He~{\sc i}~4471 & 7.0 & 1.75 & 1.92 &  -- \\
      Mg~{\sc i}~4563+71 & 3.0 & 0.15 & 0.14 &  -- \\
      $[$C~{\sc iii}$]$$+$$[$C~{\sc iv}$]$~4652 & 4.0 & 0.14 & 0.14 &  -- \\
      He~{\sc ii}~4686 & 2.0 & 0.02 & 0.01 &  -- \\
      $[$Ar~{\sc iv}$]$~4711 & 2.5 & 0.08 & 0.05 &  -- \\
      H$\beta$~4861 & 7.0 & 100.00 & 100.00&  5\\
      He~{\sc i}~4921 & 7.0 & 1.03 & 1.01 &  -- \\
      $[$O~{\sc iii}$]$~4959 & 7.0 & 20.38 & 19.35&  -- \\
      $[$O~{\sc iii}$]$~5007 & 7.0 & 64.22 & 58.69 &  5\\
      $[$N~{\sc i}$]$~5198+5200 & 5.0 & 0.0.45 & 0.41 &  -- \\
      $[$Cl~{\sc iii}$]$~5517 & 4.0 & 0.12 & 0.10 &  -- \\
      $[$Cl~{\sc iii}$]$~5537 & 5.0 & 0.73 & 0.43&  -- \\
      $[$O~{\sc i}$]$~5577 & 5.0 & 1.01 & 0.56 &  -- \\
      $[$N~{\sc ii}$]$~5755 & 7.0 & 3.11 & 2.57 &  -- \\
      He~{\sc i}~5876 & 7.0 & 13.61 & 11.00 &  -- \\
      $[$O~{\sc i}$]$~6300 & 7.0 & 3.33 & 2.56 &  -- \\
      $[$O~{\sc i}$]$~6363 & 7.0 & 0.87 & 0.67 &  -- \\
      $[$N~{\sc ii}$]$~6548 & 7.0 & 113.54 & 84.76 &  -- \\
      H$\alpha$~6563 & 7.0 & 400.44 & 297.23 &  3\\
      $[$N~{\sc ii}$]$~6584 & 7.0 & 346.77 & 253.76 &  5\\
      He~{\sc i}~6678 & 7.0 & 3.11 & 2.28 &  -- \\
      $[$S~{\sc ii}$]$~6717 & 7.0 & 12.41 & 9.11 &  -- \\
      $[$S~{\sc ii}$]$~6731 & 7.0 & 16.97 & 12.41&  5\\
      \hline
    \end{tabular}
    \label{mapintens}
    Observed H$\beta$  flux, F(H$\beta$): 1.52~$\times$~10$^{-11}$ 
erg~cm$^{-2}$~s$^{-1}$.
  \end{minipage}
\end{table}

\subsection{Robustness of the 2D flux measurements}

To check the accuracy of the technique, what follows is the comparison
between the fluxes measured directly from a single slit and those
obtained from the spectroscopic map.  Several fluxes from the slit~G
(see Figure~\ref{imhalpha}) were compared with the fluxes measured
from the region corresponding to slit~G in the maps.

Flux measurements from slit~G were obtained using the {\it splot} task
of {\sc iraf}. These measurements have been normalized to correspond to a 
region whose width is 1~arcsec (following the fact that the resolution
of the maps is 1\arcsec~pixel$^-$$^1$).  The comparison is made for 5 separate
nebular regions. The extent of each of these regions, at the position of
slit G, is given in Table~\ref{idreg}.

\begin{table}
  \centering
  \begin{minipage}{80mm}
    \caption{Nebular regions of NGC~40, as defined along slit~G.} 
    \begin{tabular}{@{}lccccc@{}}
      \hline	
      Region        		& ID  & Approximated extension\\
                    		&     & (arcsec)\\
      \hline
North Outer Shell   		& NOS & from 46 to 57\\
North Inner Rim     		& NIR & from 60 to 76\\ 
South Inner Rim     		& SIR & from 79 to 99\\
South Outer Shell   		& SOS & from 101 to 112\\
Integrated nebular emission 	& WN  & from 46 to 112\\
      \hline
    \end{tabular}
\label{idreg}
    \end{minipage}
 \end{table}

The emission-lines used for comparison were selected in such a way that 
isolated, blended, strong and weak lines were each chosen. These measurements 
are shown in Table~\ref{comparison}, together with the relative errors
between the two types of flux measurements. In more than 75$\%$ of
the cases, the discrepancies are lower than 10$\%$. In the worst
case, it is 25$\%$.

\begin{table*}
 \centering
 \begin{minipage}{154mm}
  \caption{Comparison between observed fluxes measured from the slit~G and 
from the  corresponding region in the emission-lines maps. Fluxes are given in 
erg~cm$^{-2}$~s$^{-1}$.}
  \begin{tabular}{@{}lccccccccc@{}}
  \hline
   Line Identification & \multicolumn{3}{c}{NOS} & \multicolumn{3}{c}{NIR} 
& \multicolumn{3}{c}{SIR} \\
    & slit G & map & $\delta$ ($\%$) & slit G & map & $\delta$ ($\%$) & slit G 
& map & $\delta$ ($\%$) \\
 \hline
H$\beta$ 4861 & 3.22(-14) & 3.53(-14)& 8.8 & 1.18(-13) & 1.39(-13) 
& 14.5 & 1.37(-13) & 1.41(-13) & 2.9 \\
$[$O~{\sc iii}$]$ 4959 & 3.59(-15) & 4.50(-15) & 20.2 & 2.80(-14) & 
2.86(-14) & 1.9 & 3.07(-14) & 3.24(-14) & 5.2 \\
$[$O~{\sc iii}$]$ 5007 & 1.10(-14) & 1.48(-14) & 25.2 & 8.49(-14) & 
8.85(-14) & 4.0 & 9.37(-14) & 9.97(-14) & 6.1 \\
$[$N~{\sc ii}$]$ 6548 & 3.94(-14) & 4.17(-14) & 6.5 & 1.36(-13) & 
1.60(-13) & 14.9 & 1.53(-13) & 1.59(-13) & 3.9 \\
H$\alpha$ 6563 & 1.30(-13) & 1.39(-13) & 6.2 & 4.71(-13) & 
5.47(-13) & 13.8 & 5.45(-13) & 5.57(-13) & 2.2 \\
$[$N~{\sc ii}$]$ 6584 & 1.18(-13) & 1.25(-13) & 5.6 & 4.13(-13) & 
4.85(-13) & 14.9 & 4.68(-13) & 4.78(-13) & 2.0 \\
$[$S~{\sc ii}$]$ 6717 & 5.10(-15) & 5.24(-15) & 2.7 & 1.80(-14) & 
1.90(-14) & 5.2 & 1.77(-14) & 1.68(-14) & 5.9 \\
$[$S~{\sc ii}$]$ 6731 & 6.62(-15) & 6.87(-15) & 3.6 & 2.45(-14) & 
2.82(-14) & 13.1 & 2.50(-14) & 2.45(-14) & 2.2 \\
\hline
\end{tabular}
\label{comparison}
\end{minipage}
\end{table*}

\begin{table*}
 \centering
 \begin{minipage}{127mm}
   \contcaption{}
   \begin{tabular}{@{}lccccccc@{}}
     \hline
     Line Identification & & \multicolumn{3}{c}{SOS} & \multicolumn{3}{c}{WN} 
\\
     & & slit~G & map & $\delta$ ($\%$) & slit~G & map & $\delta$ ($\%$) \\
     \hline
     H$\beta$ 4861 & & 3.00(-14) & 3.00(-14) & 0.0 & 3.88(-13) & 3.89(-13) 
     & 0.1 \\
     $[$O {\sc iii}$]$ 4959 & & 1.98(-15) & 1.84(-15) & 7.5 & 7.49(-14) & 
     7.13(-14) & 5.0 \\
     $[$O {\sc iii}$]$ 5007 & & 5.93(-15) & 6.15(-15) & 3.5 & 2.25(-13) & 
     2.21(-13) & 1.8 \\
     $[$N {\sc ii}$]$ 6548 & & 3.55(-14) & 3.57(-14) & 0.6 & 4.14(-13) & 
     4.22(-13) & 1.7 \\
     H$\alpha$ 6563 & & 1.20(-13) & 1.19(-13) & 0.3 & 1.47(-12) & 
     1.48(-12) & 0.4 \\
     $[$N {\sc ii}$]$ 6584 & & 1.08(-13) & 1.08(-13) & 0.5 & 1.28(-12) & 
     1.29(-12) & 0.6 \\
     $[$S {\sc ii}$]$ 6717 & & 4.67(-15) & 4.34(-15) & 7.4 & 5.25(-14) & 
     4.64(-14) & 13.1 \\
     $[$S {\sc ii}$]$ 6731 & & 5.58(-15) & 5.43(-15) & 2.7 & 8.27(-14) & 
     6.59(-14) & 25.4 \\
     \hline
   \end{tabular}
 \end{minipage}
\end{table*}

\begin{figure}
\begin{center}
\includegraphics[width = 0.40 \textwidth]{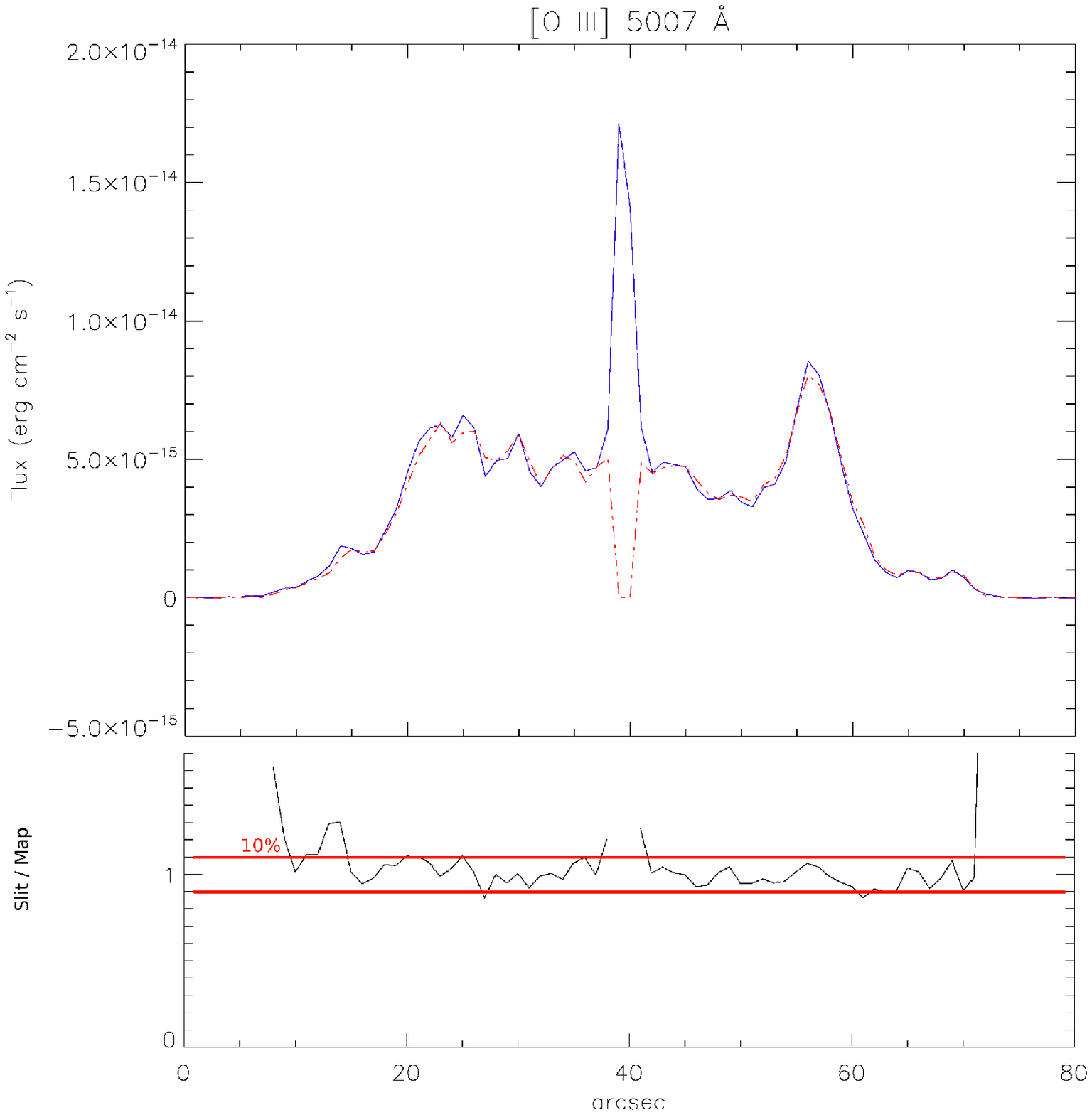}\\
\includegraphics[width = 0.40 \textwidth]{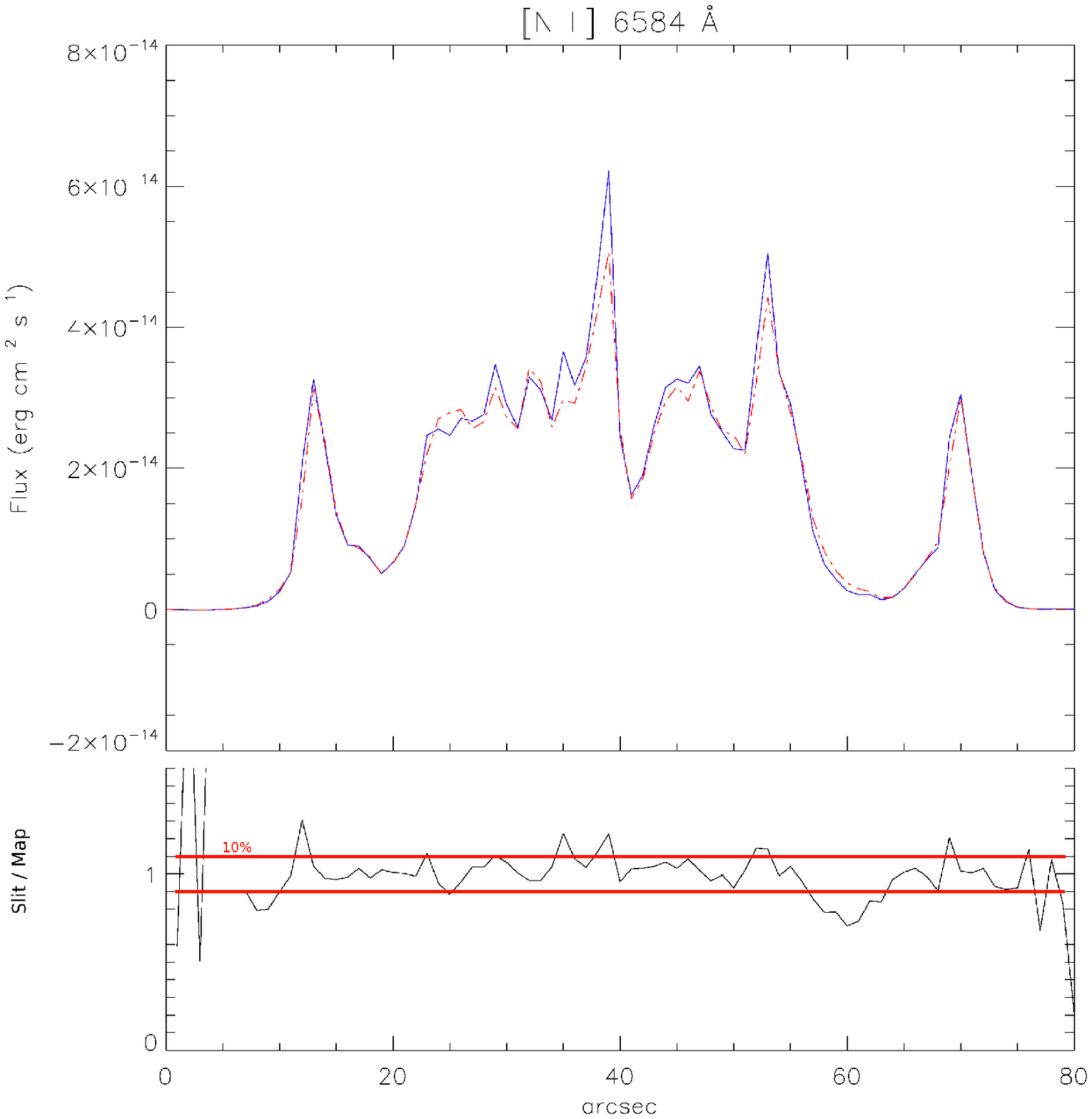}\\
\includegraphics[width = 0.40 \textwidth]{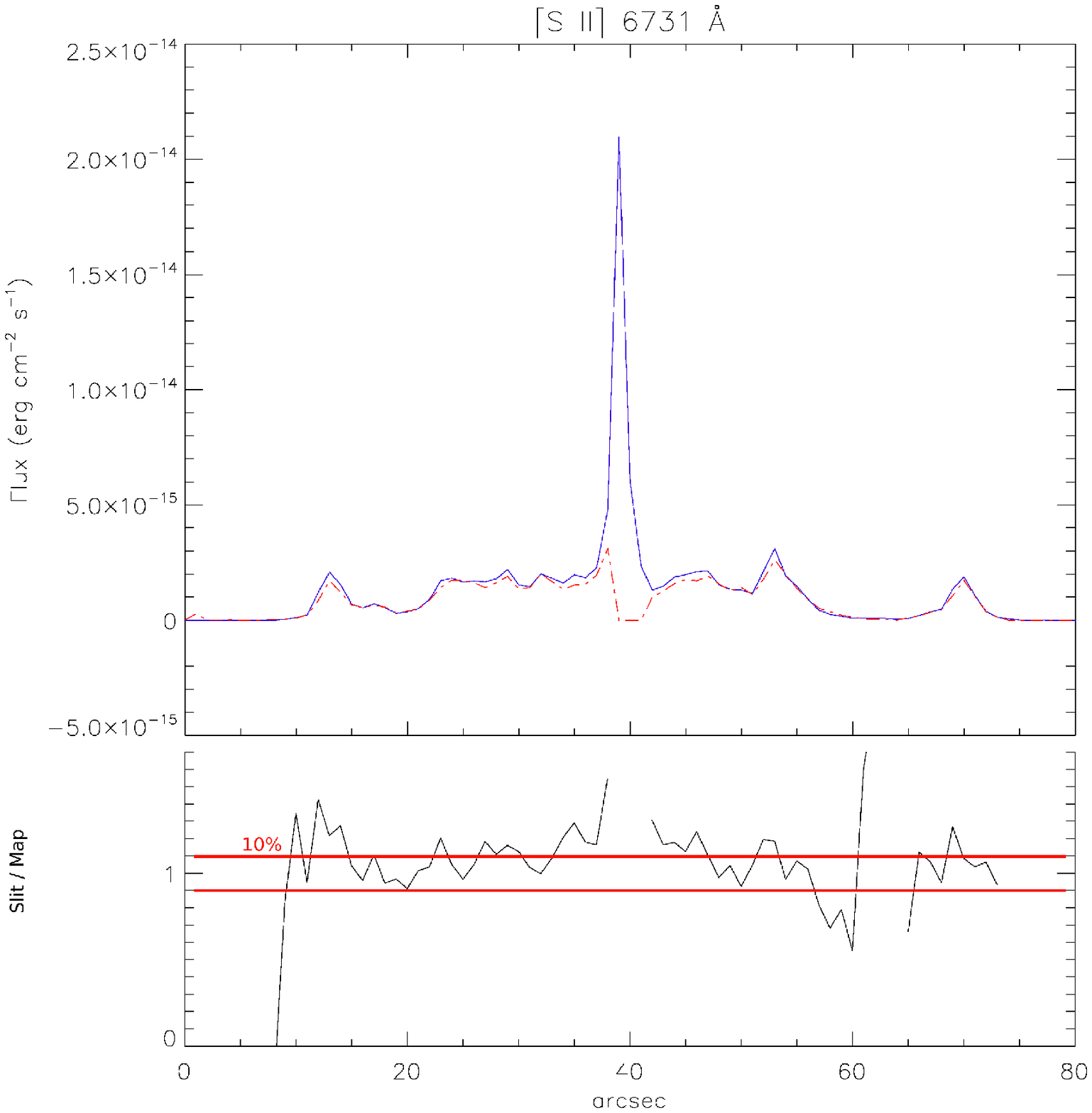}\\
\caption[]{Comparison between the slit and spectroscopic map fluxes (in 
erg~cm$^{-2}$~s$^{-1}$). From top to bottom it is shown: 
[O~{\sc iii}]~5007~\AA, [N~{\sc ii}]~6584~\AA, and [S~{\sc ii}]~6731~\AA. For 
each emission, the solid line shows the emission measured directly from 
slit~G, and the dashed line shows the emission from the spectroscopic map. In 
the bottom of each plot it is shown the ratio between both measurements. The 
region where the difference between the fluxes is lower than 10$\%$ is 
highlighted between 2 horizontal lines.}
\label{comp_flux}
\end{center}
\end{figure}

The observed discrepancies are likely due to effects specific to the regions
chosen, such as a single pixel with a discrepant value, since this value will
affect the summed flux of the region. Additionally, since the maps are 
constructed by interpolating between the slits, discrepancies might occur
due to the fact that the exact slit position is not extracted from the
map. Figure~\ref{comp_flux} shows a comparison of the spatial profile
extracted from the map and that from slit G. In this figure
 the profiles of [O~{\sc iii}]~5007~\AA, [N~{\sc ii}]~6584~\AA,
and [S~{\sc ii}]~6731~\AA ~emission lines are plotted. The profile of slit~G is
represented by the solid line, while that of the spectroscopic
map is depicted by the dashed one. Note that, with the exception of the
central region --where the central star is located--, the
correspondence between both profiles (or emission distribution) is
very good. It is also evident that the majority of the discrepant
regions correspond to either high intensity gradients or low SNR regions, in
the spatial axis. This result shows that the spectroscopic mapping
technique is, in fact, a reliable method of representing the
emission-line fluxes.

\subsection{Tools to deal with 2D emission-line maps}
To perform the usual diagnostic calculations efficiently for the
emission-line maps, it was necessary to develop a number of IDL
routines. The set of routines we built compose the {\sc 2d\_neb}
package, which is described in what follows.  The software package is divided 
in two parts: the first one is responsible for extinction correction and
the second calculates the physical and chemical conditions of the
nebula (N$_e$, T$_e$, ionic and total chemical abundances). The {\sc
  2d\_neb} package was built based on tasks of the {\sc iraf}
stsdas.analysis {\it nebular} package, adapted to work with 2D
images. As such, all the results generated by these tools are 2D
maps created in fits file format.

The routines that compose the first part of the package are based on
the \citet{b16} formalism, and are responsible for calculating the
extinction coefficient, c(H$\beta$), and correcting the data for
reddening. The Balmer lines are used to determine c(H$\beta$). For
these tasks the user can choose between H$\alpha$, H$\gamma$ or
H$\delta$ --relative to H$\beta$--, and can decide which extinction
curve will be adopted: \citet{b18} or Cardelli, Clayton, \& Mathis
(1989), with the coefficients reviewed by \citet{b11}. When using
Cardelli, Clayton \& Mathis (1989) curve, it is necessary to assume a
value for the R${_\nu}$ parametrization coefficient, defined as
R${_\nu}$~=~(A${_\nu}$~/~E(B-V)). The default value of R${_\nu}$, used
for the calculations hereafter, is 3.1 \citep{b3}. 

The remaining routines of {\sc 2d\_neb} were based on the {\sc iraf}
{\it temden} task, following the recipes given by \citet{b5}. These
routines are responsible for solving the equation of statistical
equilibrium. The atomic data used in our tools are read from {\sc
  iraf}'s atomic data directory. The electron temperature (T$_e$) and
density (N$_e$) are determined from the solution of the equations of 
statistical equilibrium. This package also calculates chemical abundances 
(ionic and total).  The latter procedures were based on the {\it ionic} and
{\it abund} tasks of the {\sc iraf} facility. To perform these
calculations, the effective recombination coefficient for the H$\beta$
and the He main transitions, not included in the {\sc iraf}'s atomic
data directory, are required. These atomic data were taken from
Benjamin, Skillman \& Smits (1999). Total chemical abundance
derivation uses the ionic correction factors (ICF) formalism of
\citet{b7}.

\subsection{{\sc 2d\_neb} package: benchmarks}
In this section we show the calculations that were performed to benchmark
the results obtained from the {\sc 2d\_neb} package. This was done by
using arbitrary input data covering reasonable PN values, and the
results of each step are checked by comparing them with the results
given by the tasks of {\sc iraf}'s {\it nebular} package.  

The first benchmark that we show concerns the performance of the extinction 
correction routine. As the {\it redcorr} task (stsdas.analysis.nebular) works
with the original coefficients from the Cardelli, Clayton \& Mathis
(1989) extinction function, we adapted our algorithm to also work with
these coefficients. We performed the extinction
correction using this extinction function and adopting
R${_\nu}$~=~3.1, as is done in {\sc iraf}. We choose 4 different
values for $\lambda$, well distributed inside the optical
spectroscopic range. For each of these wavelengths, we choose
different values of c(H$\beta$) and non-corrected observed fluxes.

Table~\ref{benchmarks1} presents the results of the extinction
correction for the tested values. The values were used to generate 2D
images that were then fed through {\sc 2d\_neb}.  Table columns are,
respectively: line identification, observed flux, extinction
coefficient and the extinction-corrected intensity given by the {\sc
  2d\_neb} (I$_{\sc 2d\_neb}$) and by {\it redcorr} (I$_{{\sc
    redcorr}}$).  Note that the last two columns are essentially equal 
(with discrepancies always lower than 1\%), indicating that our algorithm 
for deriving extinction corrections is working perfectly.

\begin{table}
  \centering
  \begin{minipage}{70mm}
    \caption{Extinction correction performance. }
    \begin{tabular}{@{}lllll@{}}
      \hline
      Line ID ($\lambda$ in \AA)& F & c(${H\beta})$ & I$_{\rm {2d\_neb}}$ 
      & I$_{{\sc redcorr}}$  \\
      \hline
      $[$O~{\sc ii}$]$~3727  & 33.4  & 1.13 & 77.2 & 77.2   \\
      $[$O~{\sc ii}$]$~3727  & 10.1  & 0.70 & 17.0 & 17.0   \\
      H$\gamma$~4101   & 14.7  & 0.26 & 16.9  & 16.9  \\
      H$\gamma$~4101   & 16.8  & 0.43 & 21.1  & 21.1  \\
      $[$O~{\sc iii}$]$~5007 & 120.8 & 0.50 & 115.7 & 115.7 \\
      $[$O~{\sc iii}$]$~5007 & 1220.0& 0.33 & 1185.5& 1185.4\\
      $[$N~{\sc ii}$]$~5755  & 2.2   & 1.20 & 1.3   & 1.3   \\ 
      $[$N~{\sc ii}$]$~5755  & 3.2   & 0.19 & 3.0   & 3.0   \\
      $[$S~{\sc ii}$]$~6731  & 20.2  & 0.38 & 15.3  & 15.3  \\
      $[$S~{\sc ii}$]$~6731  & 17.8  & 0.95 & 8.8   & 8.8   \\
      $[$O~{\sc ii}$]$~7330  & 48.3  & 0.83 & 22.5  & 22.5  \\
      $[$O~{\sc ii}$]$~7330  & 30.5  & 0.12 & 27.3  & 27.3  \\
      \hline
    \end{tabular}
    The fluxes, F, are normalized to H$\beta$~=~100.
    \label{benchmarks1}
    \end{minipage}
\end{table}

The next benchmark we performed concerns N$_e$ and T$_e$
determination.  We choose to show the results for typical density
and temperature diagnostics in photoionized nebulae, when only the
optical spectra is available. For each of these quantities we show
different values of assumed electron temperature or density, and
different values for the diagnostic line ratios.  The line ratios were
chosen in a way to cover the range of possible values given in
Osterbrock \& Ferland 2006.  \te\ and \ne\ benchmarking results are
shown in Table~\ref{benchmarks2}. In this table, columns 1 to 5 show,
respectively: the quantity (output) to be determined, the assumed \te\
or \ne, the line ratio, and the results calculated using {\sc 2d\_neb}
as well as those obtained by {\it temden} (stsdas.analysis.nebular $-$
{\sc iraf}).  Inspection of Table~\ref{benchmarks2} clearly shows
the similarity between the two results, which can also be seen in the
last column, which gives the discrepancies between the results from
{\sc 2d\_neb} and {\it temden}.  The discrepancies are, in all
cases, lower than 1$\%$.

\begin{table}
  \centering
  \begin{minipage}{78mm}
    \caption{$T_e$ and $N_e$ benchmarks.}
    \begin{tabular}{@{}lcrccr@{}}
      \hline
   Output  & Input & Ratio & {\sc 2d\_neb} & {\it temden} & $\delta$ ($\%$)\\
\hline
& $N_e$ & & $T_e$ & $T_e$ & \\
& ($cm^{-3}$) & & ($K$) & ($K$) & \\
      \hline
$T_e$$[$N~{\sc ii}$]$  &1,500  &250.00 & 7,113.9  &7,114.1 & 0.0         \\
$T_e$$[$N~{\sc ii}$]$  &2,500  &102.67 & 9,397.2  &9,399.2 & 0.0       \\
$T_e$$[$N~{\sc ii}$]$  &4,000  &70.00  & 10,809.8 &10,802.7& 0.1       \\
$T_e$$[$O~{\sc ii}$]$  &1,300  &14.59  & 14,287.5 &14,259.7& 0.2       \\
$T_e$$[$O~{\sc ii}$]$  &3,000  &21.48  & 8,271.6  &8,257.8 & 0.2         \\
$T_e$$[$O~{\sc ii}$]$  &4,300  &24.30  & 7,007.5  &7,005.9 & 0.0         \\
$T_e$$[$O~{\sc iii}$]$ &1,000  &850.00 & 7,007.5  &7,008.4 & 0.0        \\
$T_e$$[$O~{\sc iii}$]$ &3,000  &77.34  & 14,222.5 &14,223.0& 0.0        \\
$T_e$$[$O~{\sc iii}$]$ &5,000  &301.76 & 8,953.0  &8,937.7 & 0.2        \\
$T_e$$[$S~{\sc ii}$]$  &4,700  &3.73   & 9,054.8  &9,056.2 & 0.0        \\
$T_e$$[$S~{\sc ii}$]$  &2,800  &4.01   & 11,878.0 &11,897.0& 0.2       \\
$T_e$$[$S~{\sc ii}$]$  &1,300  &5.05   & 14,180.2 &14,152.4& 0.2        \\
\hline
& $T_e$ & & $N_e$ & $N_e$ &      \\
& ($K$) & & ($cm^{-3}$) & ($cm^{-3}$) & \\
\hline
$N_e$$[$O~{\sc ii}$]$  &7,000  &1.90    & 1667.9   &1,667.0 & 0.1   \\
$N_e$$[$O~{\sc ii}$]$  &9,000  &1.48    & 1,014.7  &1,014.4 & 0.0   \\
$N_e$$[$O~{\sc ii}$]$  &15,000 &2.68    & 6,036.2  &6,034.2 & 0.0    \\
$N_e$$[$S~{\sc ii}$]$  &7,500  &0.74    & 1,654.9  &1,655.3 & 0.0    \\
$N_e$$[$S~{\sc ii}$]$  &10,000 &0.87    & 1,030.8  &1,029.3 & 0.1   \\ 
$N_e$$[$S~{\sc ii}$]$  &13,500 &0.59    & 4,997.2  &4,993.7 & 0.1    \\
$N_e$$[$Cl~{\sc iii}$]$&7,000  &0.91    & 3,289.2  &3,290.5 & 0.0    \\
$N_e$$[$Cl~{\sc iii}$]$&12,000 &0.84    & 4,777.5  &4,776.7 & 0.0    \\
$N_e$$[$Cl~{\sc iii}$]$&11,000 &1.15    & 1,380.8  &1,379.4 & 0.1    \\
$N_e$$[$Ar~{\sc iv}$]$ &9,000  &0.94    & 4,952.5  &4,957.8 & 0.1    \\
$N_e$$[$Ar~{\sc iv}$]$ &10,500 &1.17    & 2,165.0  &2,164.1 & 0.0    \\
$N_e$$[$Ar~{\sc iv}$]$ &12,000 &1.28    & 1,174.4  &1,173.6 & 0.1    \\
      \hline
    \end{tabular}
    \label{benchmarks2}
    \end{minipage}
 \end{table}

 The last benchmark we show is related to ionic abundance
 determination. We choose 4 different ions, and their main wavelengths
 within the optical range, to show the functionality of our
 routines. For each ion, we assumed a fixed value of electron
 temperature and density.  Table~\ref{benchmarks3} shows the results
 of these benchmarks. The first to the sixth column shows, 
 respectively: the ion for which the abundance will be derived with
 its wavelength in \AA, the adopted electron density, the adopted
 electron temperature, and finally the ionic abundance (X/H$^+$)
 determined by {\sc 2d\_neb} and by the {\it ionic} task
 (stsdas.analysis.nebular). The last colunm shows the discrepancies,
 which are always lower than 1$\%$.

 Some other aspects of the {\sc 2d\_neb} performance were also tested, such as
 the determination of the c(H$\beta$) map, the derivation of critical
 densities, emissivities and wavelengths of each transition, and
 the calculation of the total chemical abundance maps. For space reasons, we 
 not show these benchmarks here.  
 
 As a last example of the accuracy of our
 IDL routines in {\sc 2d\_neb}, we make sure that the determination of
 c(H$\beta$) agrees with the formulation given in \citet{b16}:

\begin{equation}
c(H\beta) = \frac{log[F(\lambda)/F(H\beta)]_{teo} - 
log[F(\lambda)/F(H\beta)]_{obs}}{f(\lambda) - f(H\beta)},
\end{equation}

\noindent where [F($\lambda$)/F(H$\beta)$]$_{teo}$ is theoretical
ratio of another Balmer line relative to H$\beta$,
[F($\lambda$)/F(H$\beta$)]$_{obs}$ is the equivalent observed fluxes
ratio, and f($\lambda$) is the value of the extinction curve in a
given $\lambda$.  There is good agreement of the critical densities,
emissivities and wavelengths, of each transition, with the results
obtained from {\sc iraf}'s {\it ionic} task
(stsdas.analysis.nebular). The agreement is always equal to or better
than 99$\%$.

\begin{table}
  \begin{minipage}{71mm}
    \begin{center}
    \caption{Benchmarks of the ionic abundances. Intensities are normalized to 
    \hb~=~100.} 
    \begin{tabular}{@{}lcccccc@{}}
      \hline
      Ion / \emph{$\lambda$}(\AA) & $I$ & {\sc 2d\_neb} & {\it ionic} & 
$\delta$ ($\%$)\\
& & ($\times$10$^{-7}$) & ($\times$10$^{-7}$) & \\
      \hline
Input: & \multicolumn{2}{c}{$N_e$~=~5,000~$cm^{-3}$} & 
\multicolumn{2}{c}{$T_e$~=~13,000~$K$}\\
\hline
N$^+$    5755 &2.36  & 90.96 & 91.03 & 0.1\\
N$^+$    6548 &76.01 & 247.7  & 247.8.& 0.0\\
N$^+$    6584 &235.43& 260.2  & 260.2 & 0.0\\
      \hline
Input: & \multicolumn{2}{c}{$N_e$~=~2,000~$cm^{-3}$} & 
\multicolumn{2}{c}{$T_e$~=~8,500~$K$}\\
\hline
O$^0$    6300 &1.10  & 40.56 & 40.54 & 0.0\\
O$^0$    6363 &0.70  & 81.20 & 81.16 & 0.0\\
      \hline
Input: & \multicolumn{2}{c}{$N_e$~=~1,500~$cm^{-3}$} & 
\multicolumn{2}{c}{$T_e$~=~13,000~$K$}\\
\hline
O$^{++}$ 4959 &60.91 & 273.3  & 273.0 & 0.1\\
O$^{++}$ 5007 &112.77& 175.4  & 175.2 & 0.1\\
      \hline
Input: & \multicolumn{2}{c}{$N_e$~=~5,000~$cm^{-3}$} & 
\multicolumn{2}{c}{$T_e$~=~8,000~$K$}\\
\hline
S$^+$    6717 &2.76  &6.038 & 6.040 & 0.0\\
S$^+$    6731 &9.73  & 12.12 & 12.12 & 0.0\\
      \hline
    \end{tabular}
\label{benchmarks3}
    \end{center}
    \end{minipage}
 \end{table}

\section{Slit versus Spectroscopic map Results}
In this section we show the results (electron densities, temperatures
and ionic abundances) obtained from a single slit, and compare them
with those obtained from the  corresponding region of the spectroscopic
maps. Following the same kind of comparison that we did in Section 3.2,
slit~G will be used.  Table~\ref{slitflux} lists the
extinction-corrected intensities obtained from slit~G, for the six
nebular regions under analysis: north filament (NF), NOS, NIR, SIR,
SOS and WN (see Table~\ref{idreg} for the definition of these structures).
Table~\ref{results} lists the results obtained from these intensities
(columns 2 to 8). The electron densities and temperatures
adopted for the ionic abundance calculations were \nesii\ and
\tenii. It was not possible to determine T$_e$ and N$_e$ for the
NF. Because of this, the \nesii\ and \tenii\ values of the NES were
adopted to enable the estimation of the ionic abundances of the NF
region.

From the results of the slit G physical parameters (Table~\ref{results}), 
one can notice the very low regional variation of the \tenii. 
On the other hand, the spatial variation of the \nesii ~is significant. 
These results should be compared with those coming from the same region 
of the spectroscopic maps. The quantities shown with brackets in
Table~\ref{results} 
 ~are the  corresponding mean values from the spectroscopic map for the 
region where the slit G was placed. As we were not able to calculate 
a \teoiii ~map, the results of this parameter refer only to the slit G.

These results (those obtained directly from the slit and those obtained from 
the maps) are in agreement 
and, with the exception of the \nesii ~estimate of the NIR region (discrepancy 
of $\sim$21$\%$), all of them show discrepancy of $\sim$10$\%$ or less. 

In particular, the spatial \nesii ~variation encountered 
in the analysis of the slit~G, can be clearly visualized in the \nesii~map
(Figure~\ref{temden}). The absence of variation of \tenii, from region
to region, as shown by the slit G results, can also be visualized on the 
results obtained from the \tenii ~map. Note that the histogram of this map is
much more concentrated around the mean value than the \nesii ~histogram.

\section{Mapping Results}

The 31 emission-line maps that were constructed using the
spectroscopic mapping technique are listed in
Table~\ref{mapintens}. The  corresponding fluxes and intensities, of
the entire nebula (WN), as well as the adopted SNR cut-off values (see
Section~3.1) are also given in this table.

\begin{figure}
\begin{center}
\includegraphics[width = 0.48 \textwidth]{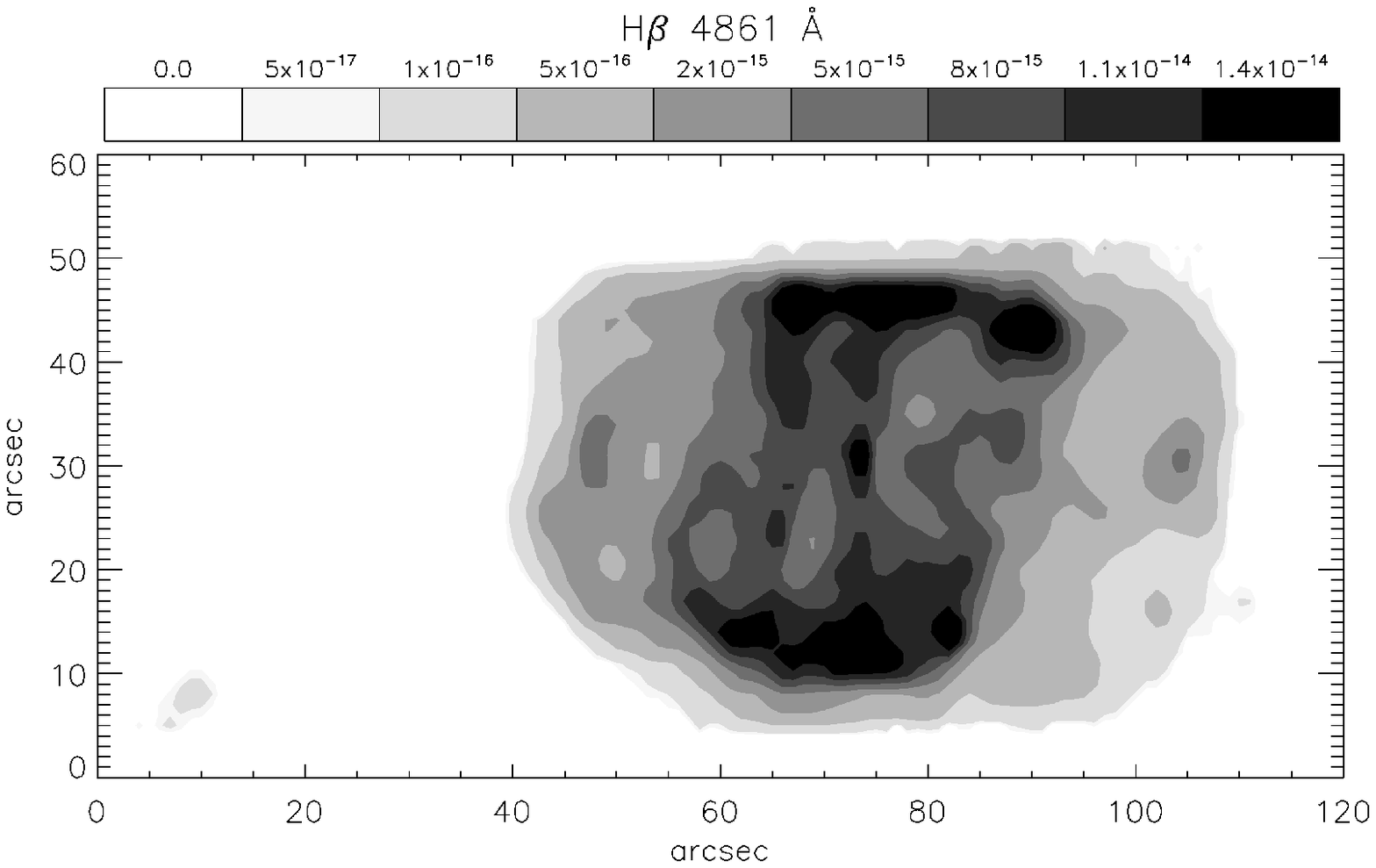}
\hfill
\includegraphics[width = 0.48 \textwidth]{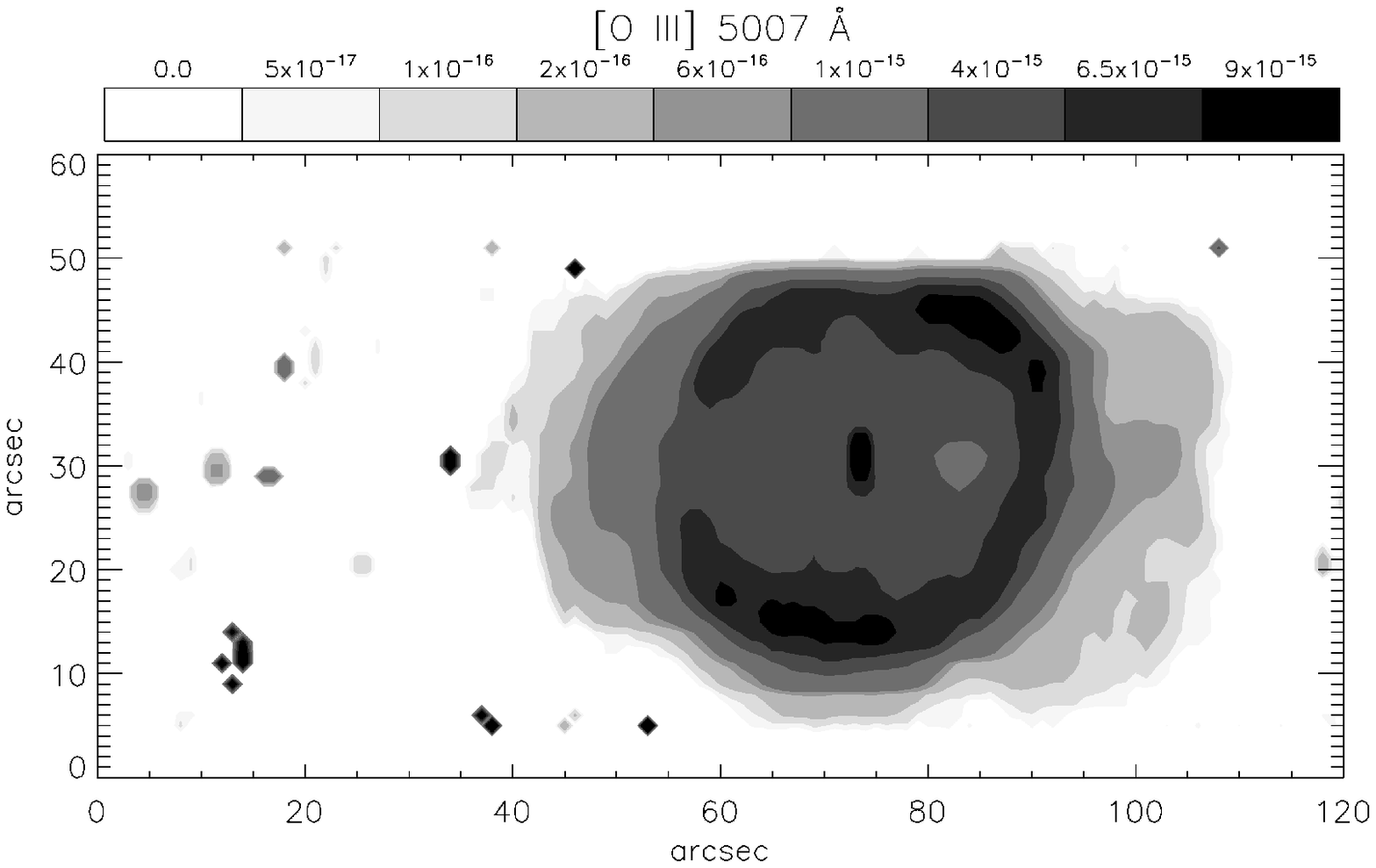}\\
\includegraphics[width = 0.48 \textwidth]{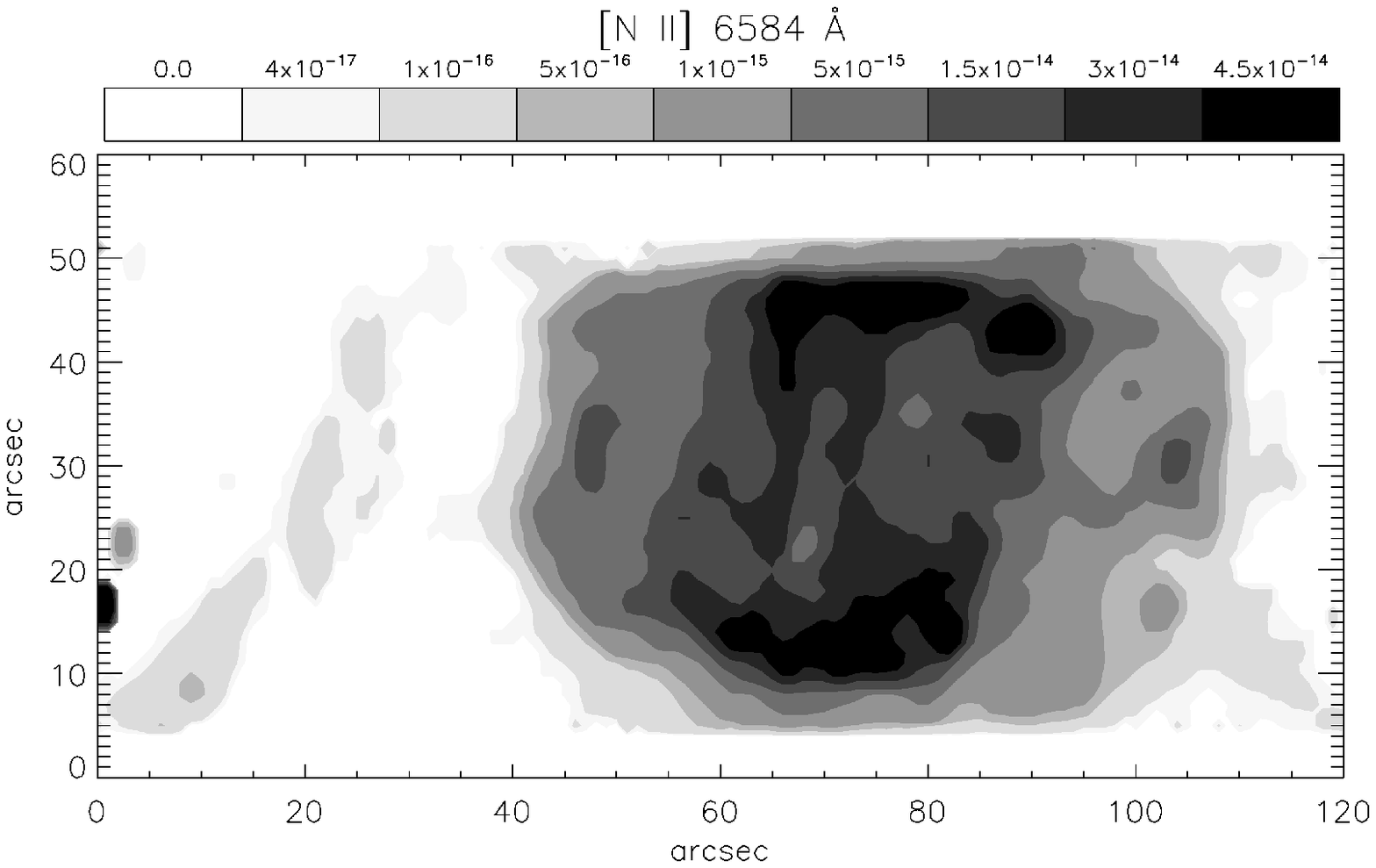}
\hfill
\includegraphics[width = 0.48 \textwidth]{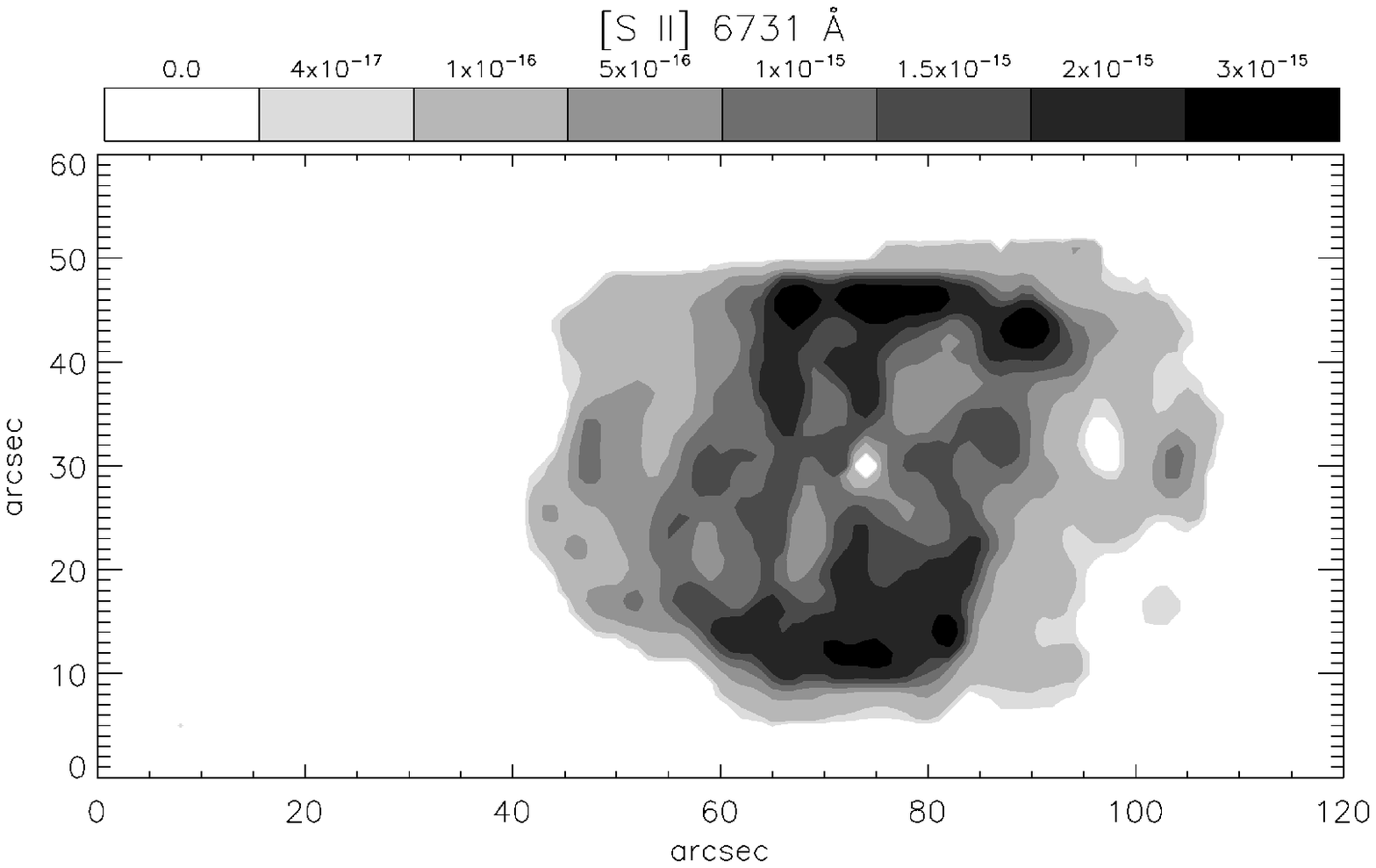}\\
\caption[]{From the top to the bottom, the following emission-line maps (in 
erg~cm$^{-2}$~s$^{-1}$) are presented: H$\beta$, 
[O~{\sc iii}]~$\lambda$~5007~\AA, [N~{\sc ii}]~$\lambda$~6584~\AA ~and 
[S~{\sc ii}]~$\lambda$~6731~\AA.}
\label{linemaps}
\end{center}
\end{figure}

The emission-line map of H$\alpha$ was already shown in 
Figure~\ref{hacontour}. Other maps --namely: H$\beta$, 
[O~{\sc iii}]~5007~\AA, [N~{\sc ii}]~6584~\AA~and [S~{\sc ii}]~6731~\AA-- 
are shown here, in Figure~\ref{linemaps}.

\begin{table*}
 \centering
 \begin{minipage}{140mm}
  \caption{Emission-line intensities of slit~G. For WN, whose emission was 
integrated along the slit, the observed flux is also given. Intensities and 
fluxes are normalized to H$\beta$ = 100.}
  \begin{tabular}{@{}lrrrrrrl@{}}
  \hline
   Line Identification & \multicolumn{7}{c}{Slit Intensities} \\
    & NF & NS & NIR & SIR & SS & \multicolumn{2}{c}{WN} \\
& & & & & & F & I \\
 \hline
H$\beta$~4861 (erg~cm$^{-2}$~s$^{-1}$) & 8.13(-16) & 2.21(-13) & 
8.18(-13) & 9.38(-13) & 2.05(-13) & 2.67(-12) \\
H10~3797 & -- & 4.80 & 4.84 & 5.68 & 5.56 & 5.34 & 6.84 \\
He~{\sc i}~3820 & -- & -- & 0.85 & 1.86 & -- & 3.11 & 3.97 \\
H9~3835 & -- & 7.45 & 7.59 & 8.87 & 7.18 & 8.49 & 10.80 \\
$[$Ne~{\sc iii}$]$~3869 & -- & -- & 1.05 & -- & -- & 1.19 & 1.50 \\
H8$+$He~{\sc i}~3888 & -- & 15.51 & 17.85 & 21.57 & 15.8 & 22.94 & 28.90 \\
H$\epsilon$$+$$[$Ne~{\sc iii}$]$$+$He~{\sc i}~3968 & -- & 15.88 & 15.39 & 16.37 & 15.71 
& 16.44 & 20.79 \\
$[$Fe~{\sc iii}$]$$+$He~{\sc i}~4009 & -- & -- & 0.18 & -- & -- & -- & -- \\
He~{\sc i}~4026 & -- & -- & 1.49 & 1.17 & -- & 1.39 & 1.70 \\
S~{\sc ii}~4033 & -- & -- & 0.33 & -- & -- & -- & -- \\
$[$S~{\sc ii}$]$$+$$[$S~{\sc ii}$]$~4069+76 & -- & 2.24 & 2.74 & 4.48 & 2.46 & 9.91 & 
12.01 \\
H$\delta$~4101 & -- & 25.28 & 24.70 & 25.54 & 24.41 & 19.34 & 23.27 \\
H$\gamma$~4340 & -- & 44.85 & 45.55 & 46.96 & 45.42 & 40.99 & 46.51 \\
$[$O~{\sc iii}$]$~4363 & -- & -- & 0.70 & 0.90 & -- & 2.66 & 3.00 \\
He~{\sc i}~4388 & -- & -- & 0.36 & 0.53 & -- & -- & -- \\
He~{\sc i}~4438$+$? & -- & -- & -- & 1.09 & -- & 9.49 & 10.49 \\
He~{\sc i}~4471 & -- & 2.67 & 2.90 & 3.18 & 2.39 & 3.89 & 4.27 \\
N~{\sc iii}~4524 & -- & -- & -- & 0.18 & -- & -- & -- \\
He~{\sc ii}~4542 & -- & -- & -- & 0.32 & -- & -- & -- \\
Mg~{\sc i}~4563+71 & -- & 1.01 & 0.51 & 0.58 & 0.75 & -- & -- \\
$[$C~{\sc iii}$]$$+$$[$C~{\sc iv}$]$~4652 & -- & -- & 2.38 & 12.27 & -- & & \\
He~{\sc ii}~4686$^*$ & -- & -- & 0.70 & 3.88 & -- & -- & -- \\
$[$Ar~{\sc iv}$]$~4711 & -- & -- & 0.51 & 0.56 & 0.78 & -- & -- \\
H$\beta$~4861 & 100.00 & 100.00 & 100.00 & 100.00 & 100.00 & 100.00 & 
100.00 \\
He~{\sc i}~4921 & -- & -- & 1.20 & 1.43 & 0.89 & 1.71 & 1.69 \\
$[$O~{\sc iii}$]$~4959 & -- & 10.91 & 23.01 & 21.98 & 6.49 & 18.76 & 18.38 \\
$[$O~{\sc iii}$]$~5007 & -- & 33.14 & 68.98 & 66.25 & 19.2 & 57.1 & 55.39 \\
$[$N~{\sc i}$]$~5198+5200 & -- & 1.37 & 0.92 & 0.93 & 1.34 & 0.33 & 0.31 \\
$[$Fe~{\sc ii}$]$$+$$[$Fe~{\sc iii}$]$~5262+70 & -- & -- & 0.41 & 0.46 & -- & 1.4 & 
1.29 \\
$[$Cl~{\sc iii}$]$~5517 & -- & 0.26 & 0.34 & 0.37 & 0.44 & 0.58 & 0.52 \\
$[$Cl~{\sc iii}$]$~5537 & -- & 0.28 & 0.37 & 0.32 & 0.47 & 0.56 & 0.50 \\
$[$O~{\sc i}$]$~5577 & -- & 0.90 & 0.41 & 0.37 & -- & -- & -- \\
$[$N~{\sc ii}$]$~5755 & -- & 2.97 & 2.92 & 2.99 & 2.92 & 3.51 & 3.02 \\
He~{\sc i}~5876 & -- & 9.60 & 11.61 & 12.38 & 9.11 & 20.62 & 17.51 \\
?~5891 & -- & -- & 0.47 & 0.47 & -- & 4.23 & 3.59 \\
$[$O~{\sc i}$]$~6300 & -- & 5.47 & 3.73 & 3.48 & 5.11 & 4.78 & 3.87 \\
$[$S~{\sc iii}$]$$+$He~II~6312 & -- & 0.54 & 0.74 & 0.69 & 0.23 & 1.03 & 0.83 \\
$[$O~{\sc i}$]$~6363 & -- & 1.78 & 1.22 & 1.19 & 1.63 & 1.42 & 1.14 \\
He~II~6406 & -- & -- & 0.12 & -- & -- & 1.16 & 0.93 \\
$[$N~{\sc ii}$]$~6548 & 110.20 & 87.57 & 84.10 & 85.19 & 88.53 & 105.90 & 
83.45 \\
H$\alpha$~6563 & 310.62 & 296.92 & 296.43 & 300.18 & 298.7 & 377.43 & 
296.95 \\
$[$N~{\sc ii}$]$~6584 & 375.66 & 267.88 & 258.05 & 256.87 & 269.22 & 328.89 & 
258.18 \\
He~{\sc i}~6678 & -- & 3.04 & 3.24 & 3.59 & 2.96 & 4.00 & 3.11 \\
$[$S~{\sc ii}$]$~6717 & 34.38 & 11.60 & 11.28 & 9.57 & 11.45 & 13.09 & 10.13 \\
$[$S~{\sc ii}$]$~6731 & 18.42 & 15.03 & 15.34 & 13.47 & 13.66 & 20.77 & 16.05 \\
\hline
\end{tabular}

$^*$ The He~II~4686 emission-line of SIR is probably blended with  
O~II 4676 emission-line.\\
We were not able to give a clear identification to the lines marked with \lq \lq ?''. 
\label{slitflux}
\end{minipage}
\end{table*}

\subsection{Extinction Correction}

\begin{figure}
\begin{center}
\includegraphics[width = 0.48 \textwidth]{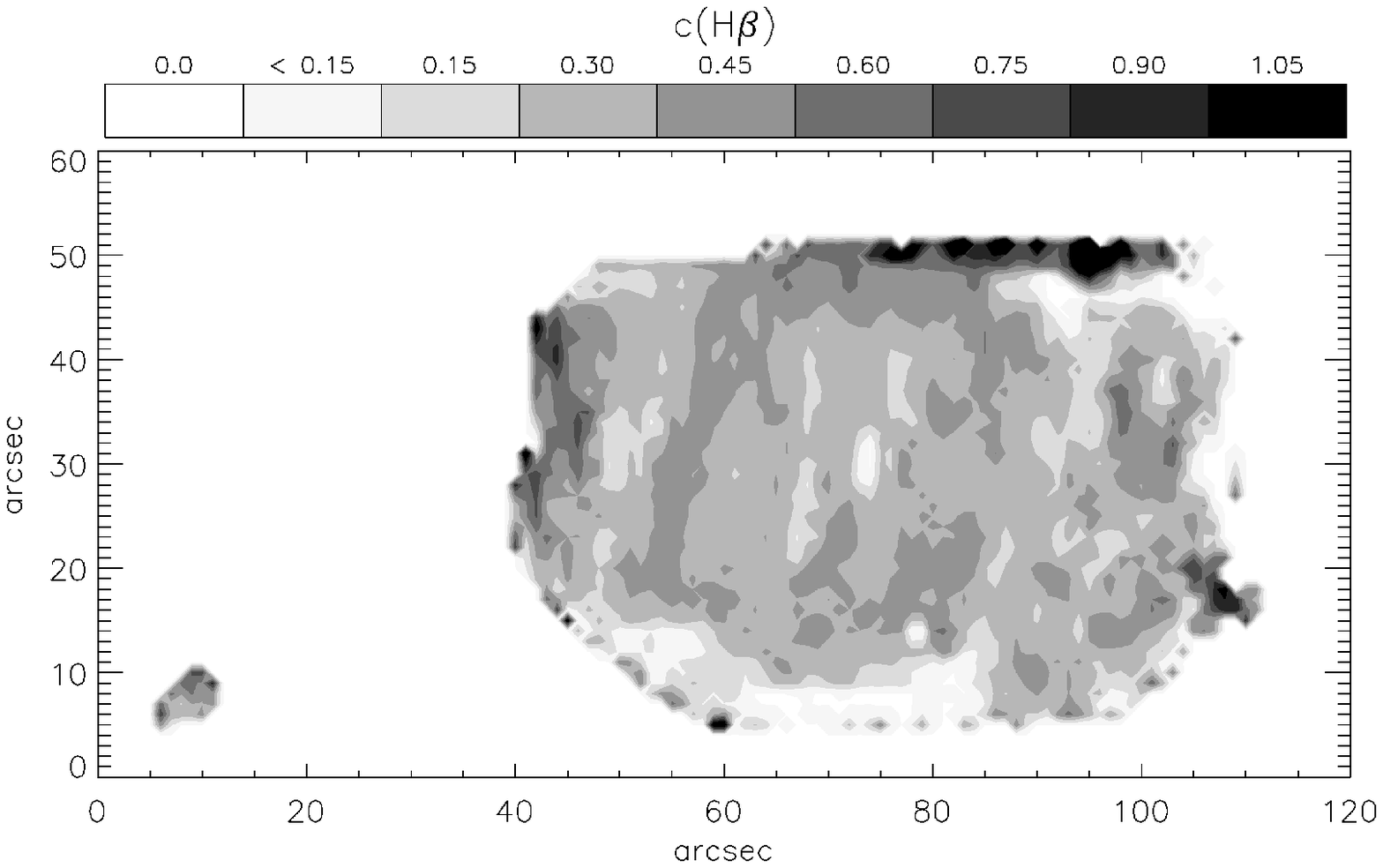} \\
\includegraphics[width = 0.48 \textwidth]{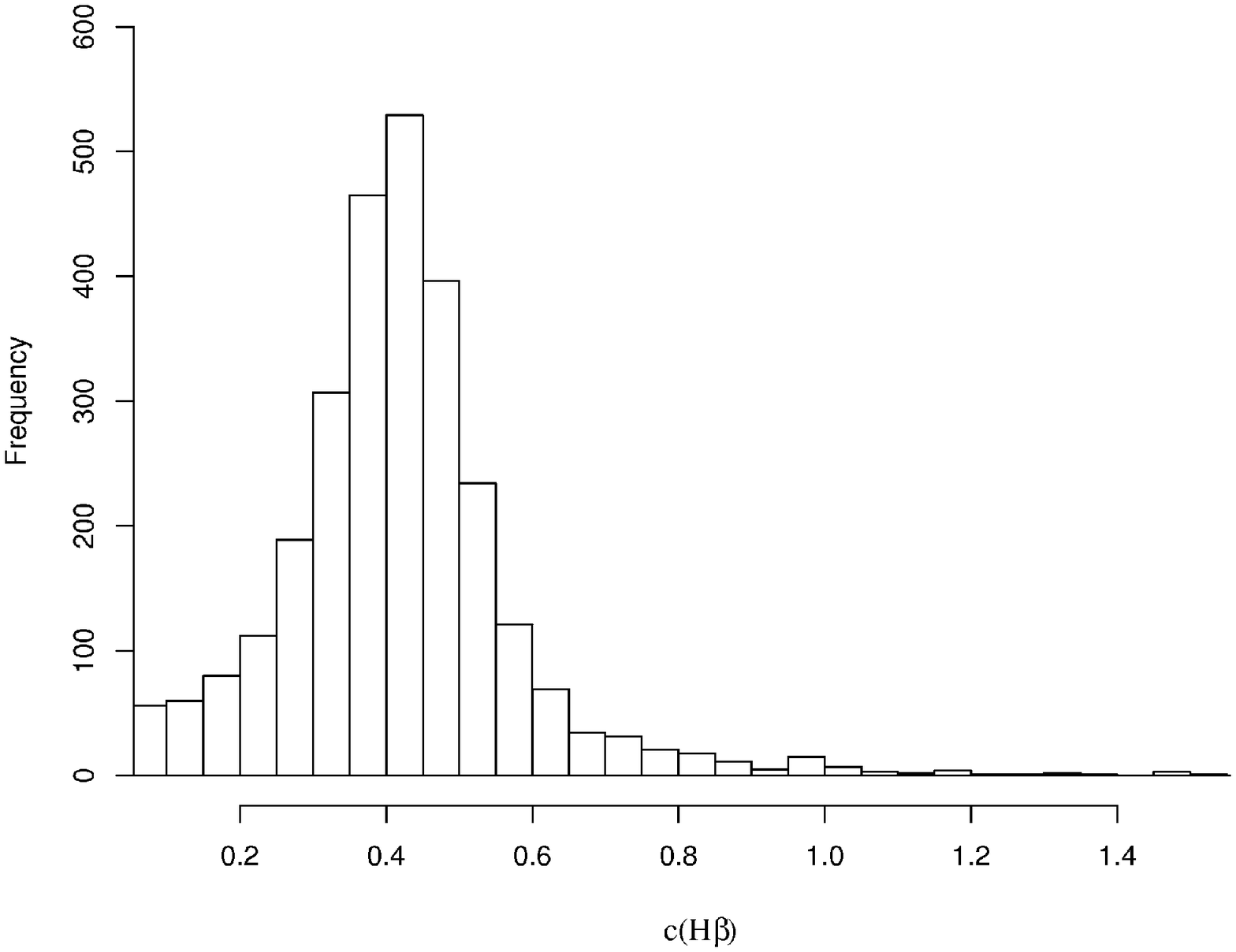}
\hfill
\caption[]{Top: c(H$\beta$) map of NGC~40. Bottom: The histogram from the 
c(H$\beta$) map. The mean value of the c(H$\beta$) map is 0.42.}
\label{cbeta}
\end{center}
\end{figure}

After creating the emission-line maps, we calculated the c(H$\beta$)
map using the {\sc 2d\_neb} package. H$\alpha$, H$\gamma$ and
H$\delta$ maps (weighted by their fluxes) and the \citet{b3}
extinction curve were all considered in deriving the c(H$\beta$) map.  The
variation of c(H$\beta$) across the nebula can be seen in
Figure~\ref{cbeta}. The spectroscopic map clearly shows that
c(H$\beta$) is not constant throughout NGC~40, and the  corresponding
histogram shows its dispersion. The mean value computed from this
map, which includes only valid values (of the pixels that survived the
noise-mask cleaning) is 0.42. The spatial variation of c(H$\beta$),
seen in Figure~\ref{cbeta}, suggests that the amount of dust across the nebula
is not constant and/or the dust grains have different
characteristics from region to region \citep{spitzer,b14}. 
Note that in studying a portion of the nebula
\citet{b4} found c(H$\beta$)$=$0.70.  In our map this region
corresponds to an area where c(H$\beta$) values are higher than average
--with minimum and maximum values of $\sim$0.22 and $\sim$1.03, respectively--
whose median, 0.62, is in agreement with \citet{b4}.

\citet{b17} and \citet{b2}, using data from various regions of the
nebula, found c(H$\beta$)$=$0.605 and c(H$\beta$)$=$0.33, respectively,
both in reasonable agreement with our map. Studying a low-ionization region 
of the nebula (not clearly identified in their paper), \citet{b1} found 
c(H$\beta$)$=$0.65. And \citet{b8}, analysing long-slit observations 
along the NGC~40 major axis found, similar to previous results, 
c(H$\beta$)$=$0.70. The previous results indicate that our mean value of 
c(H$\beta$) lies in between the values found by other authors, who used 
data from different regions of NGC~40. Furthermore, as we have access 
to the spatial variation of c(H$\beta$), it is clear why previous works 
found different extinction constants.  The trend of deriving higher 
c(H$\beta$) values in particular portions of the nebula is clearly identified 
in the map of Figure~\ref{cbeta}.

\begin{table*}
 \centering
 \begin{minipage}{168mm}
  \caption{Mean electron densities, temperatures and ionic abundances: 
slit~G versus spectroscopic mapping results.}
  \begin{tabular}{@{}lrrrrrrlrrrr@{}}
  \hline
  \hline
   Parameter & \multicolumn{6}{c}{Slit G results} & \multicolumn{3}{c}{Map results} & \multicolumn{2}{c}{Literature} \\
    & NF & NS & NIR & SIR & SS & WN & & & & & \\
 \hline
& & & & & & & & Mean & Inf.L.& Pottasch$^1$ & Liu$^2$ \\
\hline
\nesii ~(cm$^-$$^3$) & -- & 1,500 & 1,750 & 2,050 & 1,100 & 1,750 & & 1,650 & -- & 2,100 & 1,750\\
& & (1,350) & (2,250) & (2,200) & (1,150) & (1,850) & & & & & \\
\tenii ~(K) & -- & 9,050 & 9,100 & 9,100 & 9,050 & 9,050 & & 8,850 & -- & 7,500 & 8,400\\
 & & (8,550) & (8,800) & (8,750) & (8,850) & (8,800) & & & & & \\
\teoiii ~(K) & -- & -- & 11,600 & 12,950 & -- & 11,900 &  & -- & -- & 10,500 & 10,600\\
\\
He$^{+}$/H$^{+}$ 4471 & -- & -- & -- &-- & --&-- & & 4.86(-2) & 1.98(-2) & 4.8(-2) & 6.1(-2)\\
He$^{+}$/H$^{+}$ 5876 &-- &-- &-- &-- &-- &-- & & 7.45(-2) & 5.94(-2) & 4.4(-2) & 6.2(-2)\\
He$^{+}$/H$^{+}$ 6678 &-- &-- &-- &-- &-- &-- & & 6.11(-2) & 3.70(-2) & -- & 5.9(-2)\\
He$^{+}$/H$^{+}$ Mean &--&--&--&--&--&--&& 7.08(-2) & 5.66(-2) &--& 6.2(-2)\\
He$^{++}$/H$^{+}$ 4686 &-- &-- & --&-- &-- &-- & & 1.32(-3) & 4.09(-5) & -- & 3.4(-5)\\
He/H    &--&--&--&--&-- & -- & & 7.08(-2) & 5.66(-2) & $>$ 4.6(-2) &  -- \\
He/H$_{(ICF - P)}$ & -- & -- & -- & -- & -- & -- & &  9.32(-2) &  7.45(-2) & --  &  -- \\
He/H$_{(ICF - L)}$ & -- & -- & -- & -- & -- & -- & &  1.18(-1) &  9.40(-2) &  --  &  1.19(-1)\\
\\
N$^{0}$/H$^{+}$ 5198$+$5200&--&1.81(-6)&1.35(-6) & 1.41(-6) & 1.77(-6) & 3.69(-6) & & -- & -- & -- & --\\
N$^{+}$/H$^{+}$ 5755&--&6.95(-5)& 6.65(-5) & 6.61(-5) & 6.98(-5) & 6.57(-5) & & 7.32(-5) & 4.52(-5) & -- & --\\
N$^{+}$/H$^{+}$ 6548&8.53(-5)&6.77(-5)&6.46(-5)&6.50(-5)&6.83(-5) & 6.46(-5) && 6.92(-5) & 6.90(-5) & -- & 7.58(-5)\\
N$^{+}$/H$^{+}$ 6583&9.86(-5)&7.02(-5)&6.72(-5)&6.64(-5)&7.04(-5) & 6.73(-5) && 7.05(-5) & 7.05(-5)& 1.01(-4) & --\\
N$^{+}$/H$^{+}$ Mean & 9.56(-5) &6.96(-5)& 6.65(-5) & 6.60(-5)& 6.99(-5)& 6.66(-5)&& 7.05(-5) & 7.04(-5) & -- & --\\
\\
O$^{0}$/H$^{+}$ 5577 & -- & 1.53(-4) & 6.83(-5) & 6.00(-5) & -- & 6.71(-5) & & -- & -- & -- & --\\
O$^{0}$/H$^{+}$ 6300&--&1.56(-5) & 1.05(-5) & 9.63(-6) & 1.46(-5) & 1.11(-5) & & 9.98(-6) & 6.37(-6)& -- & --\\
O$^{0}$/H$^{+}$ 6363&--&1.59(-5) & 1.07(-5) & 1.02(-5) & 1.46(-5) & 1.25(-5) & & 1.03(-5) & 3.99(-6) & -- & --\\
O$^{0}$/H$^{+}$ Mean && 2.95(-5) & 1.45(-5) & 1.31(-5) & 1.46(-5) & 1.49(-5) && 1.02(-5) & 6.51(-6) &-- & --\\
O$^{++}$/H$^{+}$4363 & -- & -- & 7.43(-5) & 9.28(-5) & -- & 7.14(-5) & & -- & -- & --\\
O$^{++}$/H$^{+}$4959&--&1.58(-5)& 3.29(-5) & 3.10(-5) & 9.44(-6) & 2.87(-5) && 3.55(-5) & 3.12(-5) & -- & 1.20(-5)\\
O$^{++}$/H$^{+}$ 5007&--&1.56(-5)& 3.42(-5) & 3.24(-5) & 9.69(-6) & 3.00(-5) && 3.21(-5) & 3.08(-5) & 1.9(-5) & --\\
O$^{++}$/H$^{+}$ Mean &--& 1.57(-5) & 3.41(-5) & 3.25(-5) & 9.63(-6) & 3.00(-5)&& 3.43(-5) & 3.31(-5) &-- & --\\
O/H &--& 4.51(-5) & 4.86(-5) & 4.56(-5) & 2.43(-5) & 4.49(-5) && 4.10(-5) & 3.97(-5) & 5.3(-4)& 8.83(-3) \\
\\
Ne$^{++}$/H$^{+}$ 3869 & -- & -- & 1.75(-6) & -- & -- & 1.68(-6) & & -- & -- & 8.3(-7) & 5.33(-7)\\
\\
S$^{+}$/H$^{+}$ 4069 & -- & 8.21(-7) & -- & -- & 9.76(-7) & 3.63(-6) & & -- & -- & -- & 1.17(-6)\\
S$^{+}$/H$^{+}$ 6717&2.99(-6)&1.01(-6)&1.03(-6)&9.21(-7)&9.01(-7) & 9.86(-7) && 8.88(-7) & 7.63(-7) & -- & 1.24(-6)\\
S$^{+}$/H$^{+}$ 6731&1.24(-6)&1.01(-6)&1.04(-6)&9.20(-7)&9.03(-7) & 9.90(-7) && 8.83(-7) & 7.53(-7) & 1.48(-6) & --\\
S$^{+}$/H$^{+}$ Mean & 2.38(-6) &1.00(-6)& 1.03(-6) & 9.21(-7) & 9.02(-7)& 1.18(-6)&& 8.85(-7) & 7.74(-7) && \\
S$^{++}$/H$^{+}$ 6312& -- & 2.01(-6) & 2.70(-6) & 2.47(-6) & 8.66(-7) & 2.69(-6) & & -- & --& 2.8(-6) & 1.34(-6)\\
\\
Cl$^{++}$/H$^{+}$ 5517&--&4.38(-8)&5.79(-8) & 6.37(-8) & 7.17(-8) & 5.18(-8) && 1.02(-7) & 1.88(-8) & 5.9(-8) & 4.19(-8)\\
Cl$^{++}$/H$^{+}$ 5537&--& 5.31(-8)&6.87(-8)& 5.82(-8) & 9.11(-8) & 5.28(-8) && 4.04(-7) & 1.86(-7) & 7.5(-8) & -- \\
Cl$^{++}$/H$^{+}$ Mean &&4.87(-8)& 6.35(-8) & 6.12(-8) & 8.16(-8)& 5.23(-8)&& 3.99(-7) & 1.91(-7) & -- & --\\
\\
Ar$^{+3}$/H$^{+}$ 4711&--&--&1.15(-7)&1.27(-7)&1.74(-7)& 1.22(-7) &&3.92(-7) & 4.94(-8) & -- & --\\
\hline
\hline
\end{tabular}
\label{results}
Pottasch$^1$: \citet{b17}. \\
Liu$^2$: \cite{b8,b9}. \\
**The central star was not considered in the determination of the physical and chemical parameters 
of whole nebula, WN. \\
\end{minipage}
\end{table*}

The c(H$\beta$) map was used for extinction correction of the 31
emission-line maps. In this process, each pixel of each spectroscopic map 
had its extinction correction done based on the  value of its corresponding 
pixel in the c(H$\beta$) map. This procedure was performed by {\sc 2d\_neb}.

After the extinction correction we calculated the mean intensities for
each emission-line of NGC~40 by integrating the emission of each
spectroscopic map. These mean intensities were normalized to
H$\beta$~=~100 and are shown in the last column of
Table~\ref{mapintens}, Section 3.1. Note that, as a consequence of
using the c(H$\beta$) map to correct each emission-line map, if a
given line-map pixel has survived the noise-mask cleaning in its map,
but has not survived this process in the c(H$\beta$) map, then the 
pixel is not a valid pixel in the subsequent extinction-corrected 
emission-line map. As such the mean intensities, shown in the last 
column of Table~\ref{mapintens}, may not correspond to the same number 
of pixels as the mean fluxes (shown in the third column of the same table).

\subsection{\ne\ and \te\ Physical Properties}

Using the {\sc 2d\_neb} the \tenii ~and \nesii ~maps were
calculated. For \nesii, ~we first assumed a constant electron
temperature, equal to 9,000~K, throughout the map. The resultant map
was subsequently used to calculate the \tenii ~map, and then the \nesii
~was recalculated using this \tenii ~map.

\begin{figure}
\begin{center}
\includegraphics[width = 0.48 \textwidth]{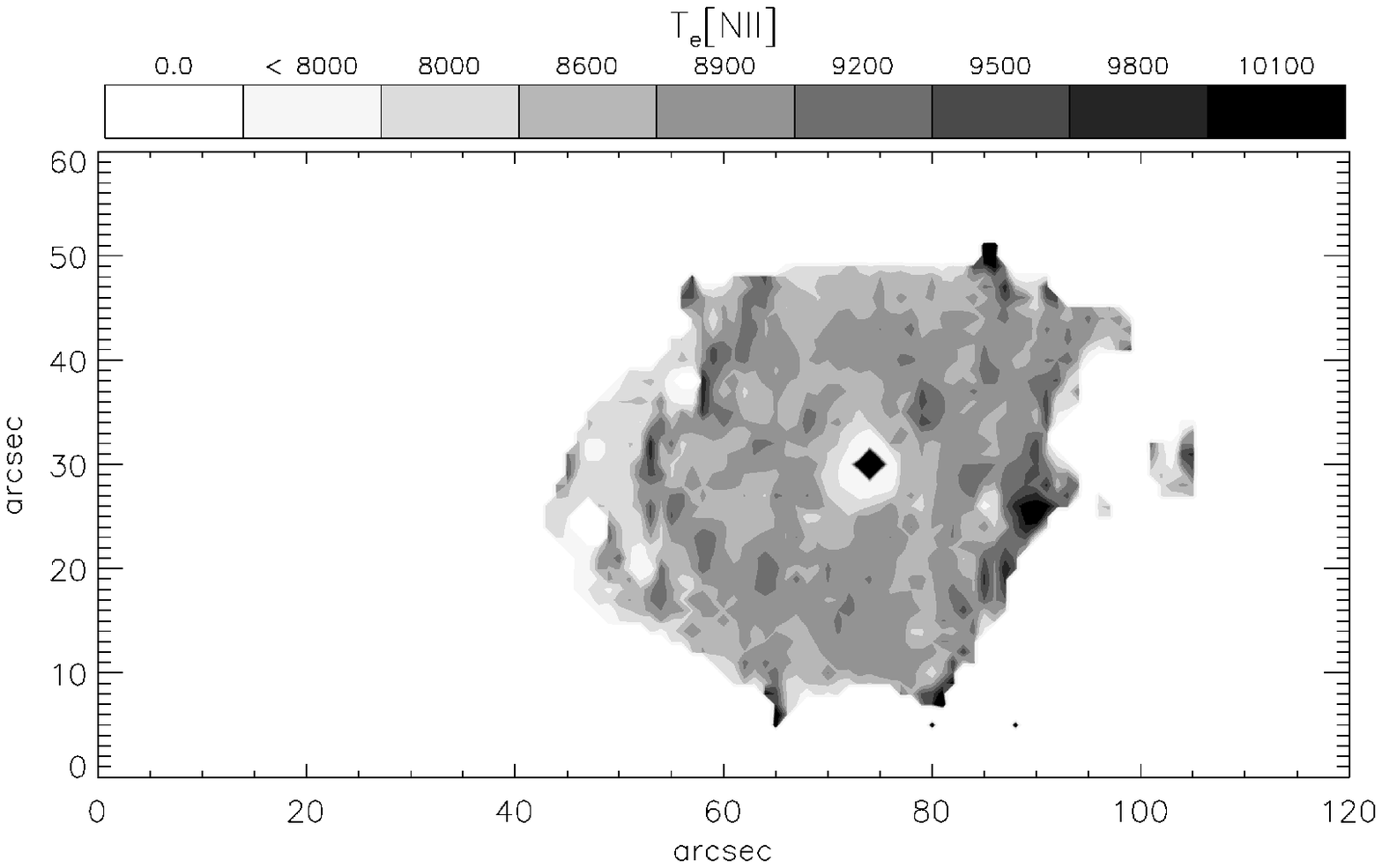}
\hfill
\includegraphics[width = 0.48 \textwidth]{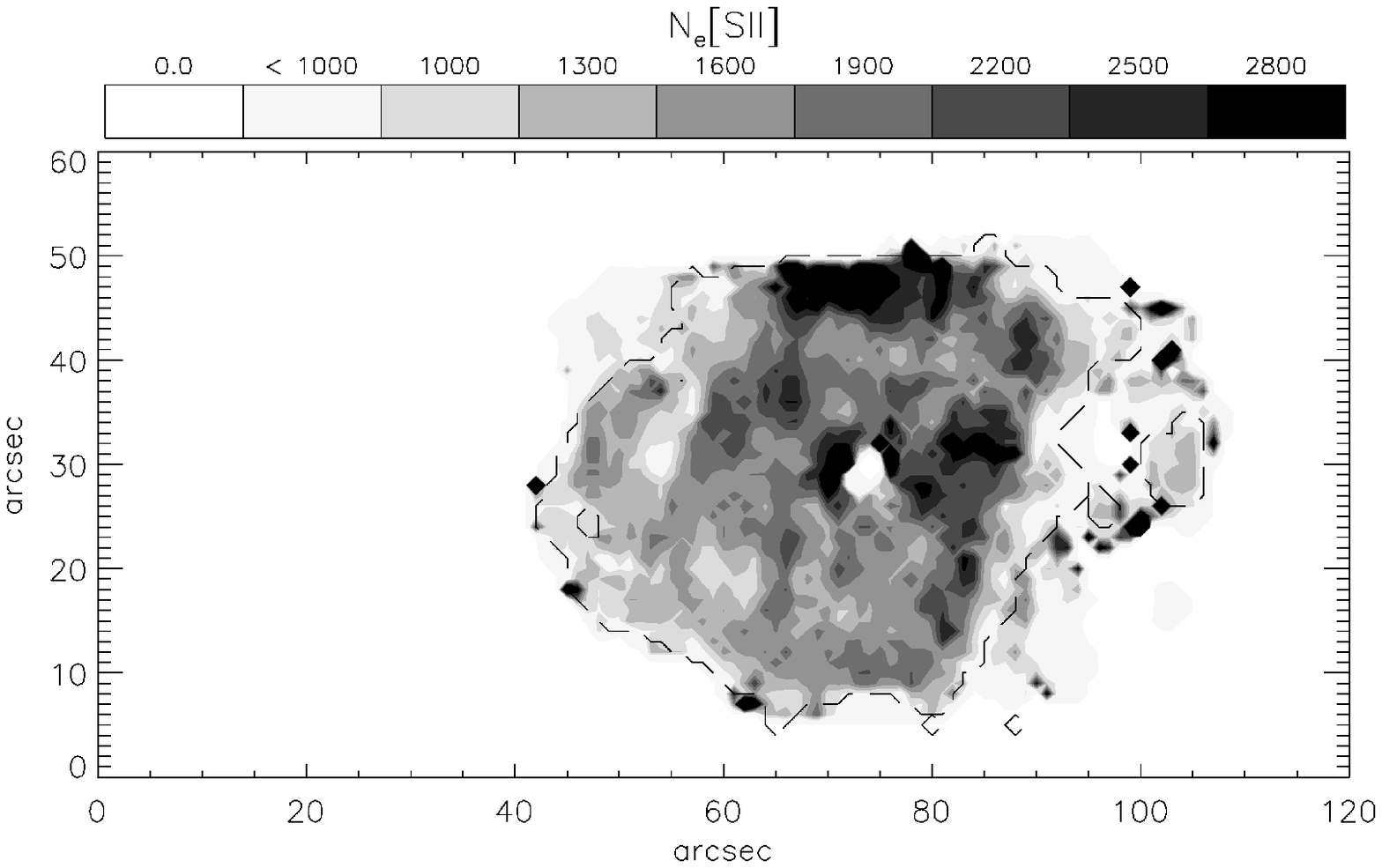}
\hfill
\includegraphics[width = 0.23 \textwidth]{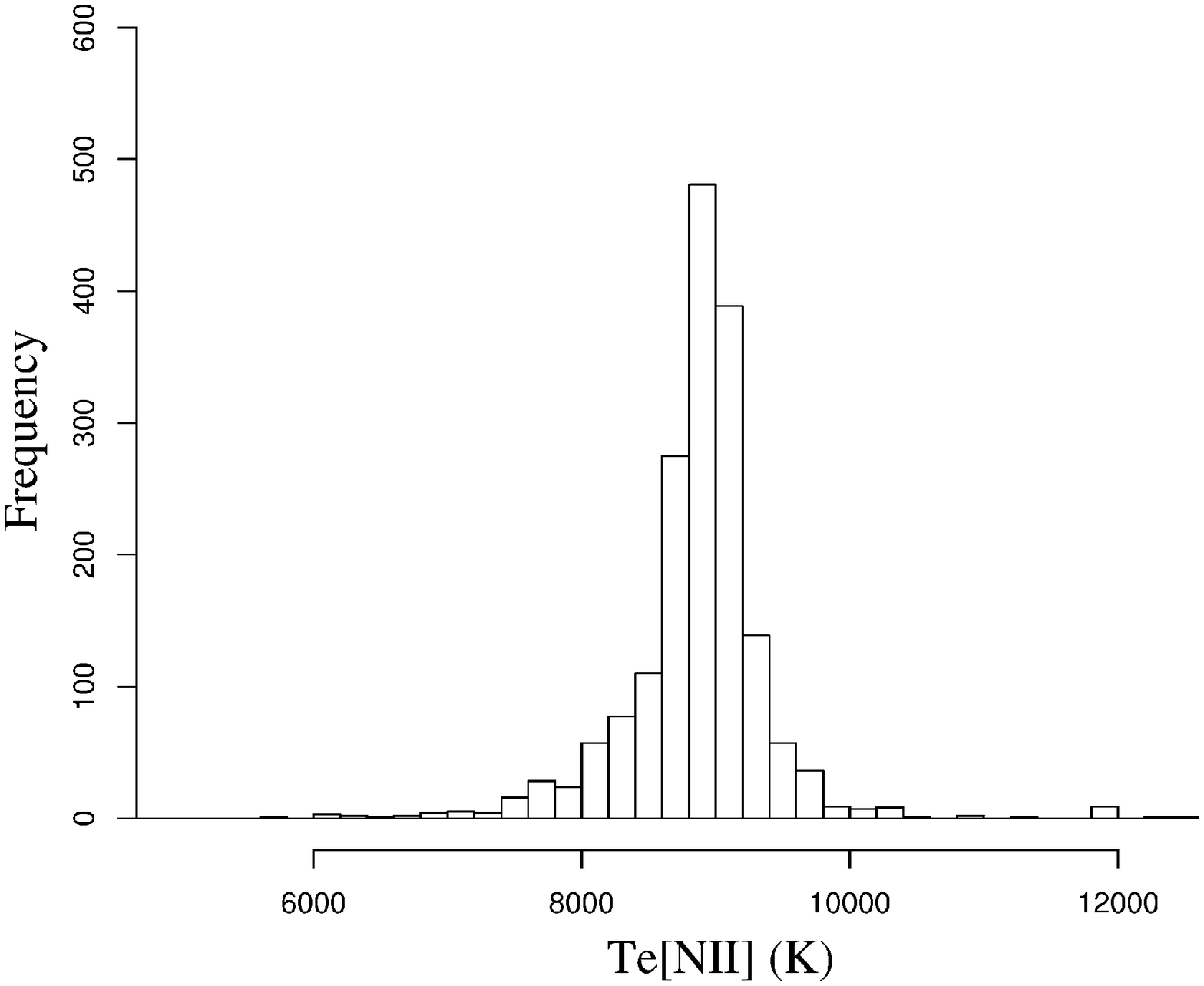}
\includegraphics[width = 0.23 \textwidth]{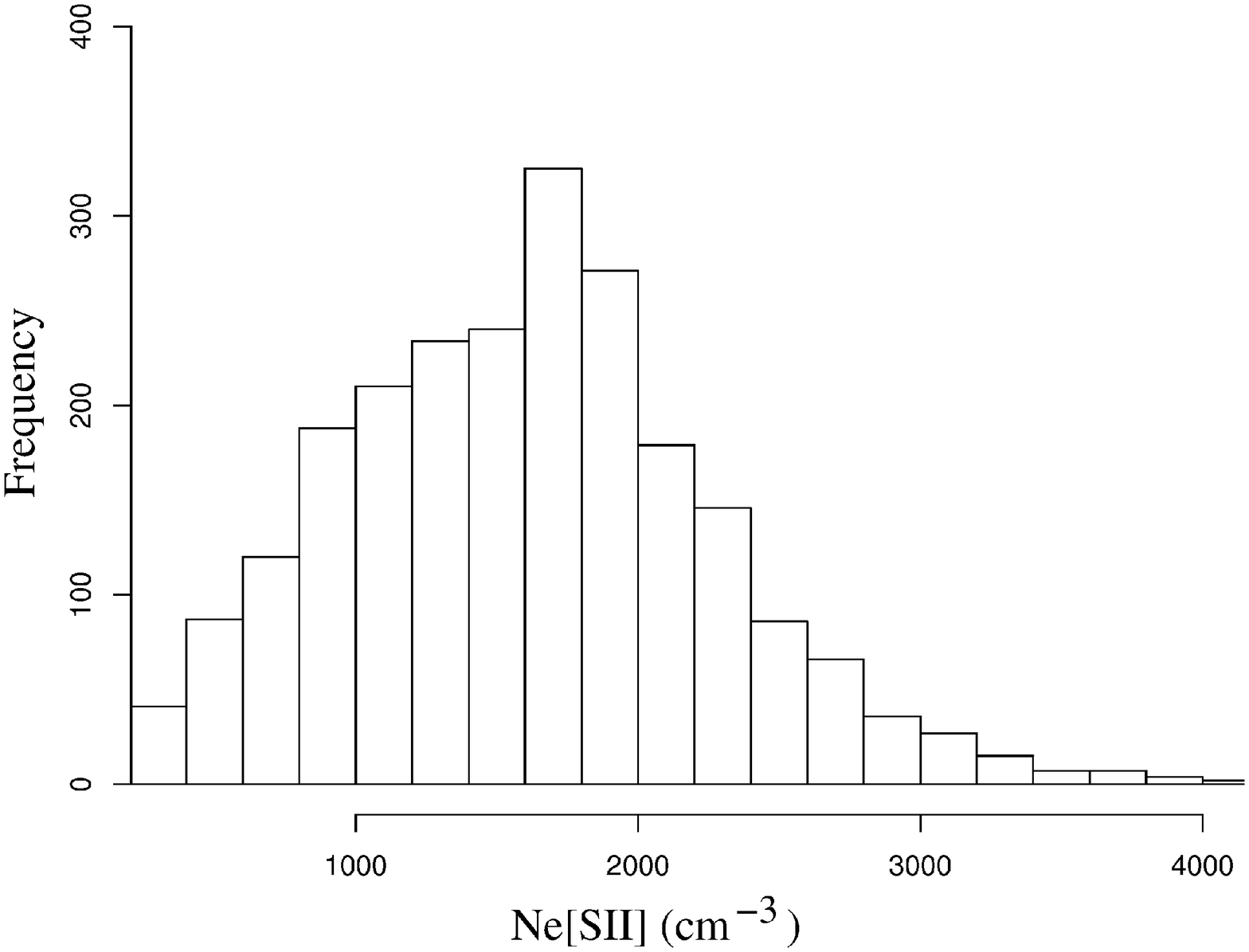}\\
\caption[]{From the top to the bottom, the following NGC~40 physical properties plots are 
shown: \tenii ~map, (in K), \nesii ~map (in cm$^{-3}$),  \tenii 
~histogram and \nesii ~histogram. The region inside the dashed line, 
on the \nesii ~map, was calculated adopting the electronic temperature 
from the \tenii ~map. A temperature equal of 9,000~K has been adopted 
for the outer region.}
\label{temden}
\end{center}
\end{figure}

The region with valid pixels in the \tenii ~map was limited by the
noise-mask cleaning of the [N~{\sc ii}]~5755~\AA ~emission-line map.
As the determination of \nesii ~depends on the electron temperature,
we assumed a constant temperature equal to 9,000~K for the non-valid
pixels of the \tenii ~map. \tenii ~and \nesii ~maps are shown in
Figure \ref{temden}. In the \nesii ~map, the dashed line indicates the
outer limit between the region in which \nesii\ has been calculated by
adopting \tenii~=~9,000~K, and the region in which the \tenii ~map was
used.

The mean values of \tenii ~and \nesii ~are, respectively, 8,850~K and
1,650~cm$^-$$^3$.  Comparing with the literature (Table~\ref{results};
columns 10 and 11), the mean values we found for \nesii ~and
\tenii\ are marginally in agreement with previous determinations of 
the physical properties in NGC~40. Notice that the \nesii ~value determined 
by \citet{b17} (2,100~cm$^{-3}$) is very similar to the values that we found 
for the NIR and SIR regions, when we used the data from the slit~G (and from 
the \nesii ~map, at that same region). On the other hand, one can notice that 
the \nesii ~map shows great variations, even 
if we consider the internal region only. The values found by \citet{b8} 
(1,750~cm$^{-3}$) and by \citet{b6} (1,800~cm$^{-3}$) approach our mean \nesii 
~map result. They both used the same line intensities. The \tenii ~results 
from \citet{b8} (8,400~K) and \citet{b6} (8,600~K) again are closer to 
the mean value of the \tenii ~map than the one found by \citet{b17}. 
Actually, when we look at the \tenii ~map, we hardly see any region
with values as low as that found by \citet{b17} (7,500~K), and most of the 
nebula shows \tenii~$>$~8,000~K.

It was not possible to calculate the \teoiii ~spectroscopic map due to
the very faint emission of [O~{\sc iii}] 4363~\AA. In cases like this,
the usual analysis made using a single slit has an advantage since the
emission is integrated throughout a region, or throughout the whole
nebula, thus the very faint emission-line flux is improved in signal to
noise.  This procedure may allow the calculation \teoiii. In fact, in
the case of NGC~40, we did derive \teoiii in two nebular
regions (NIR and SIR) using the fluxes of slit~G, as shown in
Table~\ref{results}.

\subsection{Ionic and Total Chemical Abundances}

\begin{figure}
\begin{center}
\includegraphics[width = 0.48 \textwidth]{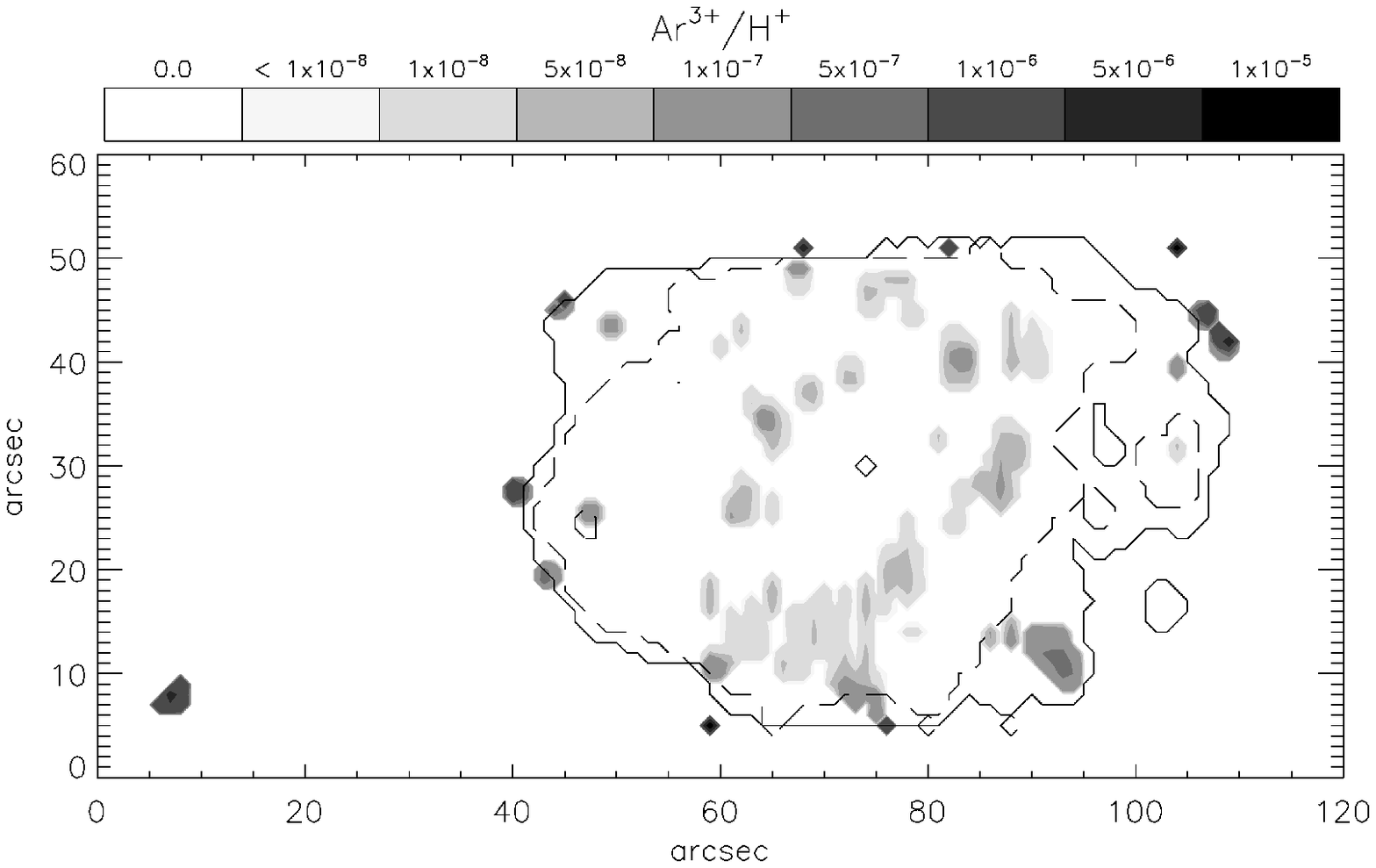}\\
\includegraphics[width = 0.48 \textwidth]{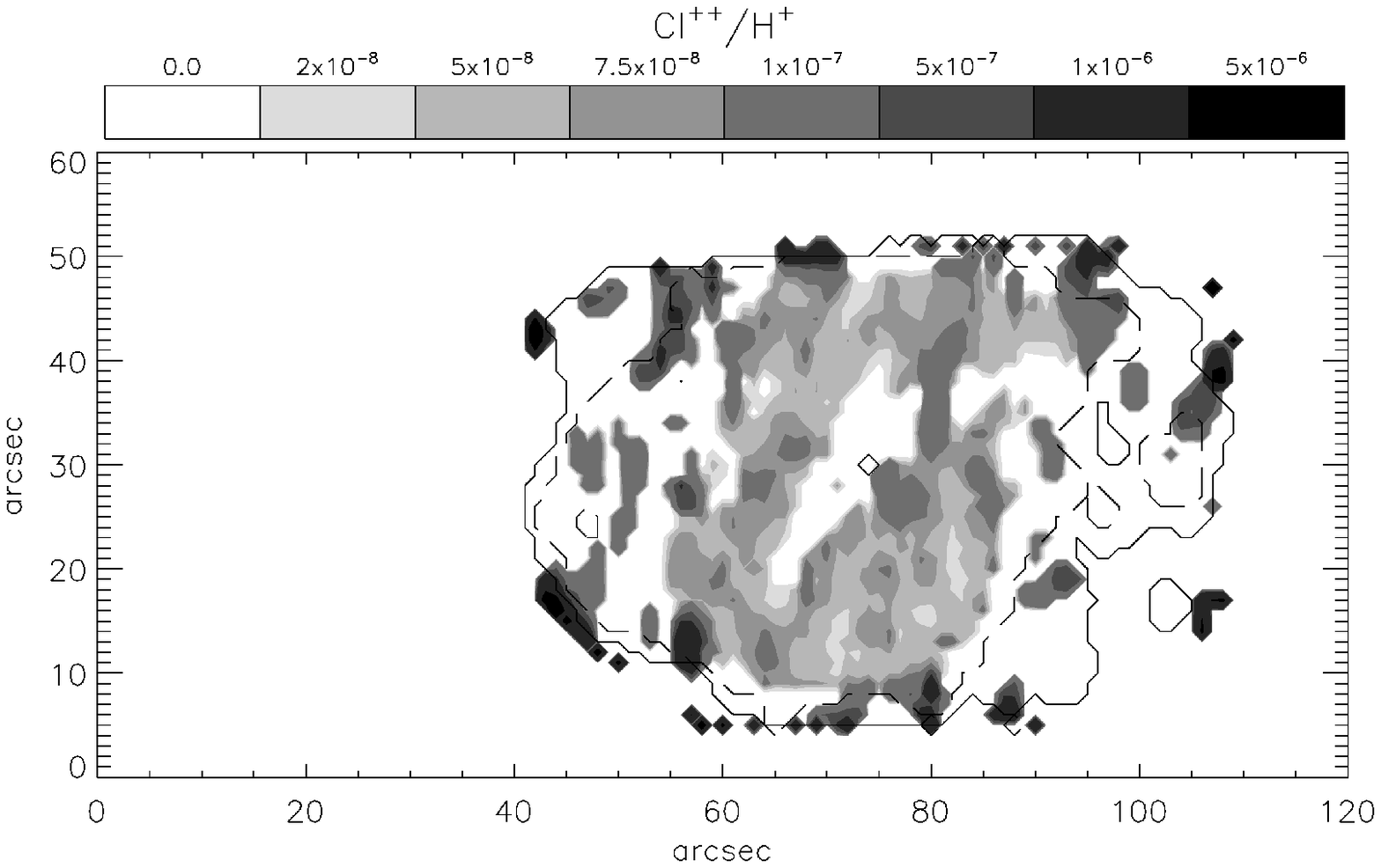}\\
\includegraphics[width = 0.48 \textwidth]{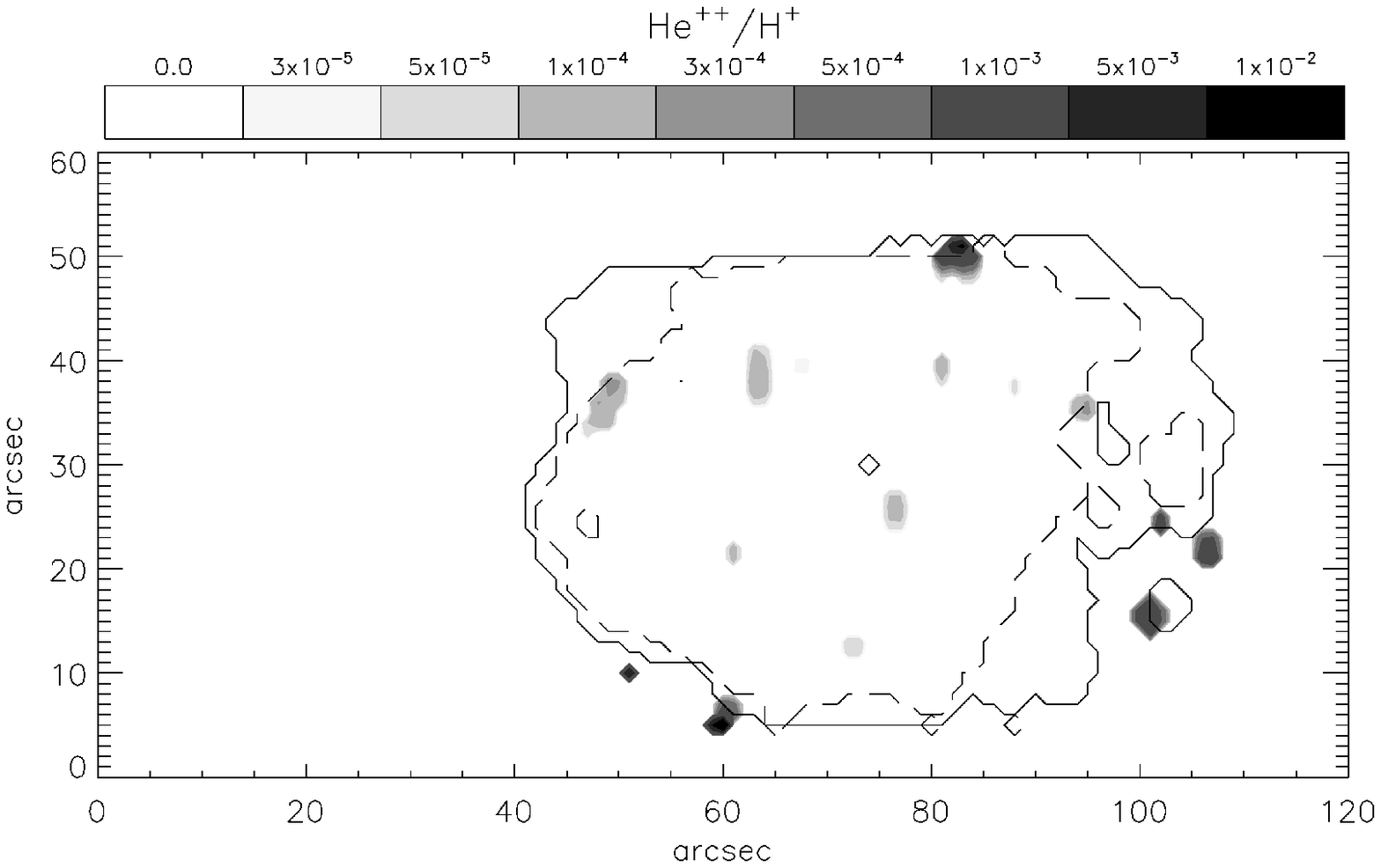}\\
\includegraphics[width = 0.48 \textwidth]{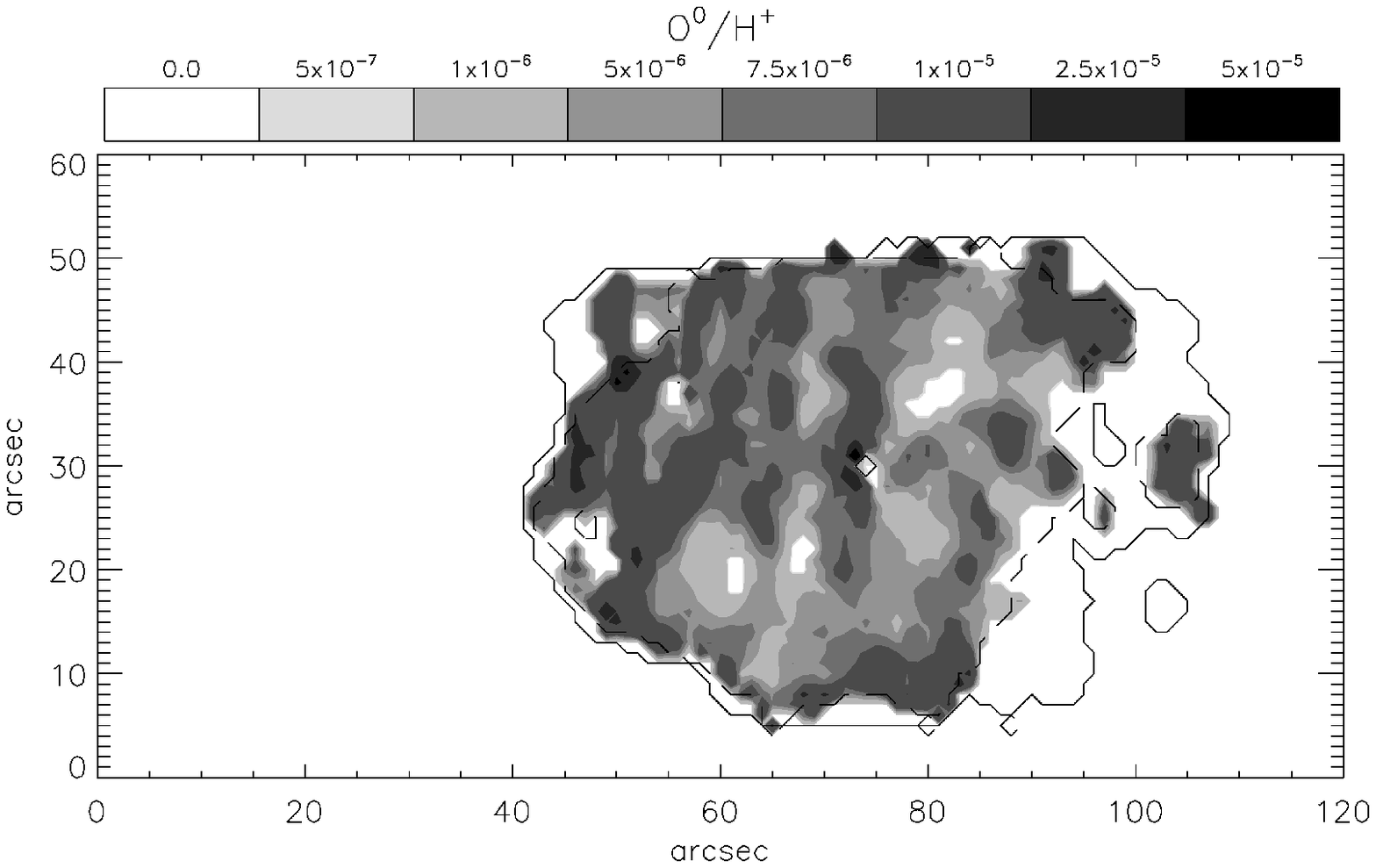}\\
\caption[]{From the top to the bottom, the following ionic abundance maps 
are presented: Ar~{\sc iv}, Cl~{\sc iii}, He~{\sc ii} and O~{\sc i}.}
\label{ionic1}
\end{center}
\end{figure}

\begin{figure}
\begin{center}
\includegraphics[width = 0.48 \textwidth]{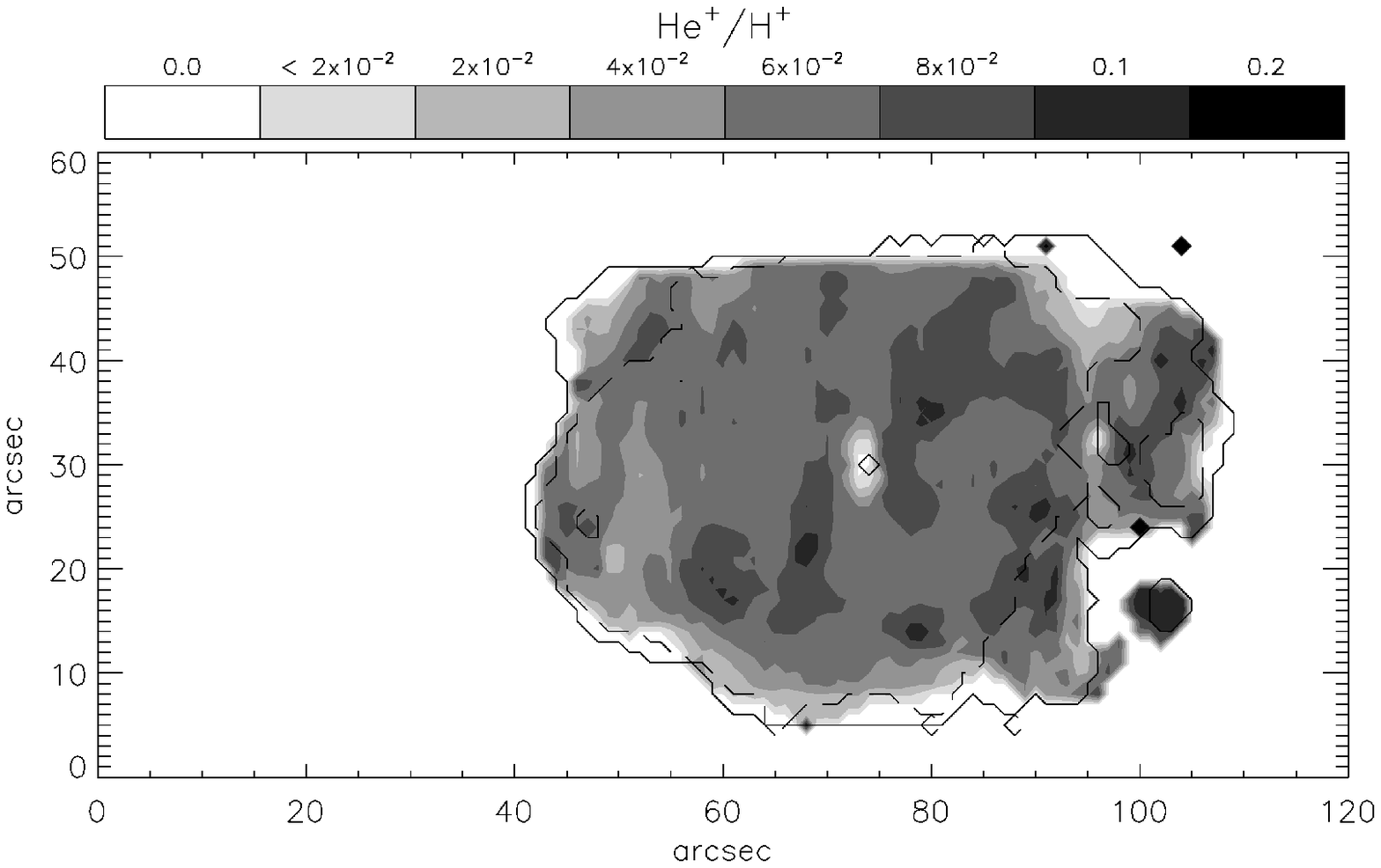}
\hfill
\includegraphics[width = 0.48 \textwidth]{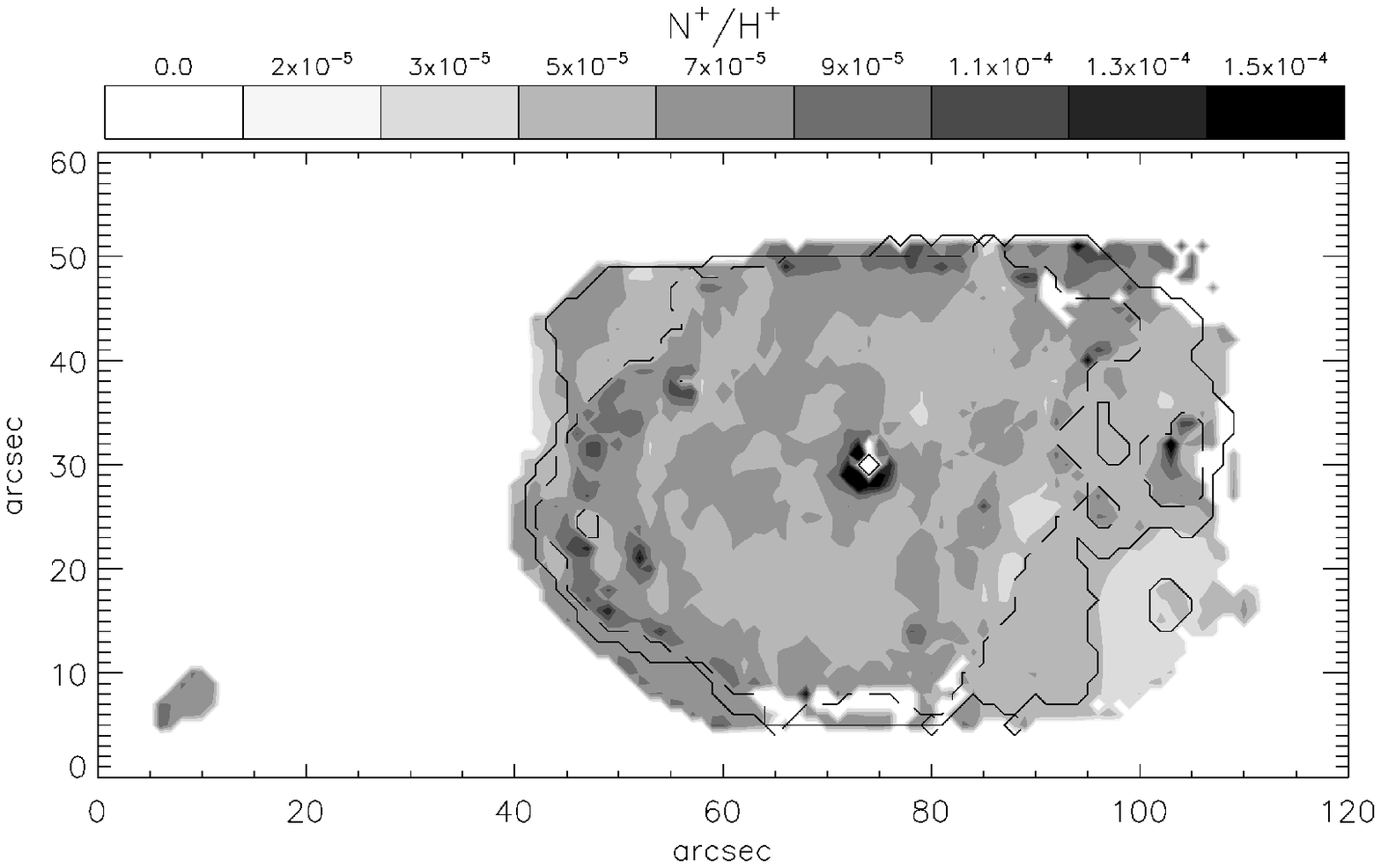}\\
\includegraphics[width = 0.48 \textwidth]{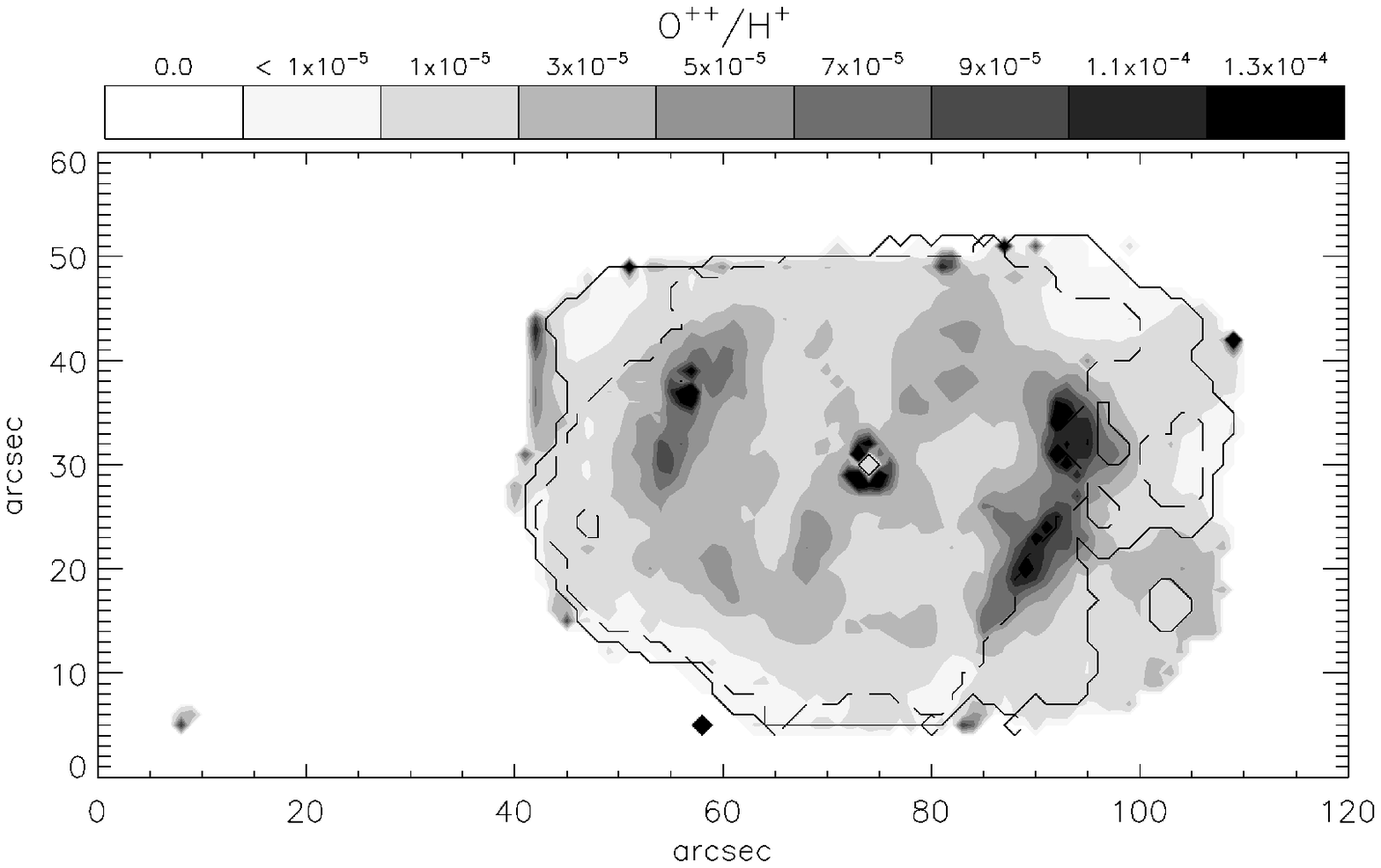}
\hfill
\includegraphics[width = 0.48 \textwidth]{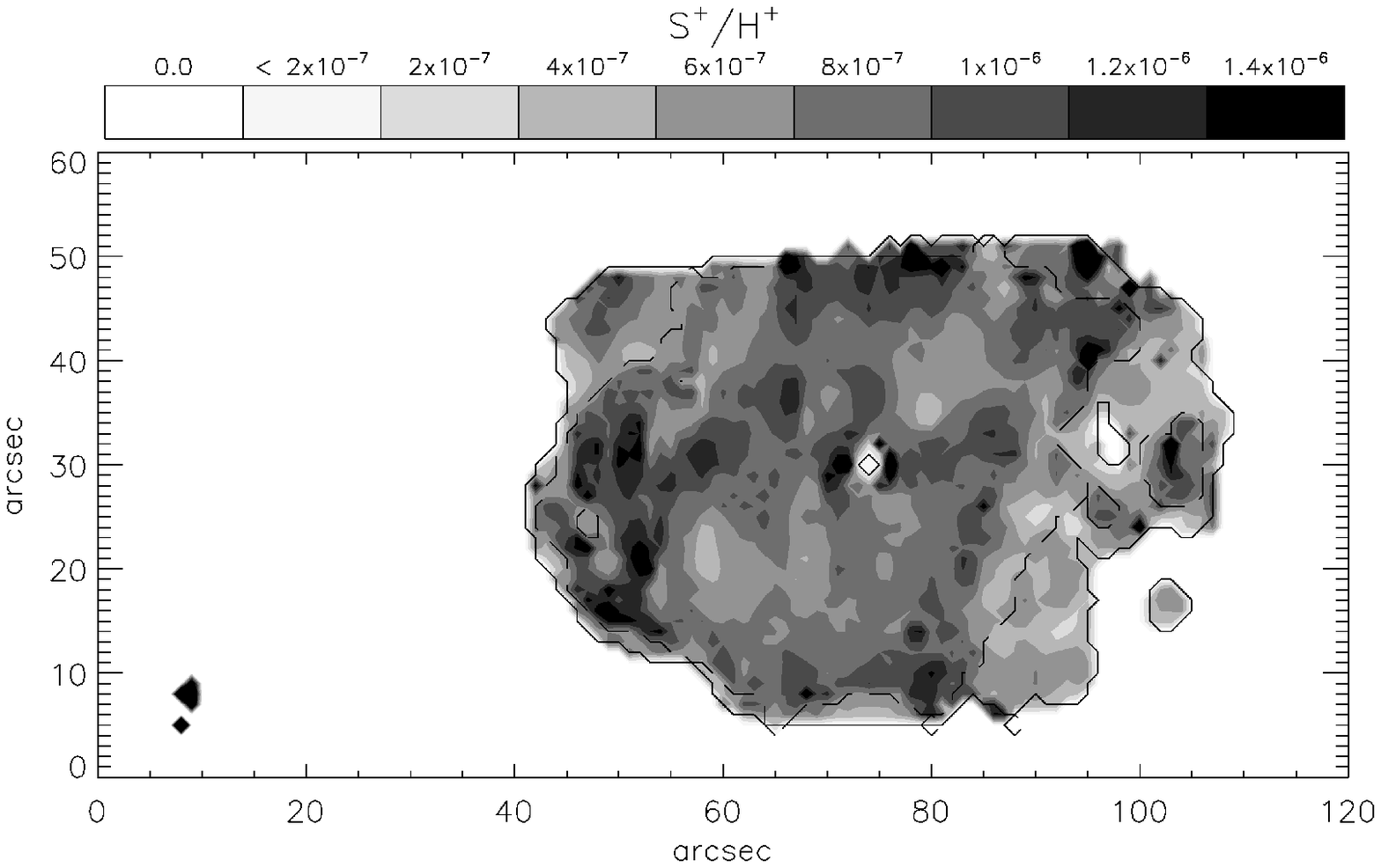}\\
\caption[]{From the top to the bottom, the following ionic abundance maps 
are presented: He~{\sc i}, N~{\sc ii}, O~{\sc iii} and S~{\sc ii}.}
\label{ionic2}
\end{center}
\end{figure}

\tenii ~and \nesii ~maps were used to calculate the ionic abundance
maps.  For the non-valid pixels of these maps we assumed constant values, 
T$_e$~=~9,000~K and N$_e$~=~1,600~cm$^{-3}$. When more than one
emission-line was available for determining the ionic abundance of a
certain ion, all the strong lines were used, each resulting in an
independent ionic abundance map. Then, the ionic abundance map was generated 
from these individual maps weighted by intensity.  These are the 
ionic maps shown in Figure~\ref{ionic1} and Figure~\ref{ionic2}.

The mean values of the ionic abundances are listed in
Table~\ref{results}, in the first column of the map results.  We draw
attention to the fact that these mean values were calculated only over
the pixels that have survived the noise-mask cleaning. This means that
the abundance maps which have non-valid pixels inside the nebular
regions whould have, in reality, lower mean abundances if the entire
nebula were considered.

As an attempt to determine the abundances representative of
the whole nebula (WN, in the tables) for the cases that have 
non-valid pixels inside the nebular region, we calculated, for each abundance 
map, the sum of the ionic abundances of all pixels and divided the result by the
number of pixels that compose the nebula\footnote{Since all abundance maps
are restricted by the locations where the c(H$\beta$) map is valid, we
used the number of valid pixels of the c(H$\beta$) map for the number
of pixels that compose the nebula}. These results represent the lower 
limits of the  corresponding NGC~40 abundances. These lower limits 
(Inf.L.) are also shown in Table~\ref{results}, column~9.

The mean ionic abundances determined in this paper can be compared
with the results found in the literature, by inspecting
Table~\ref{results}. For some cases the lower limit (column 9) is
closer to the previous ionic determinations. That is obvious for
He$^{++}$/H$^{+}$~$\lambda 4686$ abundance, whose map shows almost no
presence of that specie. The ionic abundances from
Cl$^{++}$/H$^{+}$~$\lambda 5517$, Cl$^{++}$/H$^{+}$~$\lambda 5537$,
N$^{+}$/H$^{+}$~$\lambda 6583$, O$^{++}$/H$^{+}$~$\lambda 4959$,
O$^{++}$/H$^{+}$~$\lambda 5007$, S$^{+}$/H$^{+}$~$\lambda 6717$ and
S$^{+}$/H$^{+}$~$\lambda 6731$ show a somewhat different result when
compared with the literature. The discrepancies vary from 70$\%$ up
to a factor of 3. These discrepancies can be due to the fact that
here, we are analysing a bigger area of the object: the whole nebula,
instead of one long slit, as in the case of \cite{b8}, and a set of
small portions, in the case of \cite{b17}).  Notice, for instance, that
despite the fact that the mean value of our Cl$^{++}$/H$^{+}$ map
(3.99~$\times$~10$^{-7}$) is higher than the mean value from \citet{b8}
(4.19~$\times$~10$^{-8}$) and \citet{b17} (5.9~$\times$~10$^{-8}$ and
7.5~$\times$~10$^{-8}$), the inner rim presents a lower
Cl$^{++}$/H$^{+}$ abundance when compared to the shell (see
Figure~\ref{ionic1}). The result we found for the inner rim is closer
to the results from the literature.

Examining the ionic distributions of O$^{0}$/$^{+}$H$^{+}$, N$^{+}$/H$^{+}$, 
S$^{+}$/H$^{+}$, He$^{+}$/H$^{+}$ and that of O$^{++}$/H$^{+}$, in 
Figure~\ref{ionic1} and Figure~\ref{ionic2}, we see a region (a diagonal, from 
left to right, that includes the location of the central star) in which ionic
fractions coming from the low-ionization species (O$^{+}$/H$^{+}$, N$^{+}$/H$^{+}$
and S$^{+}$/H$^{+}$) are less abundant when contrasted with the neighbor-nebular
portions. On the other hand, the map of O$^{++}$/H$^{+}$ and
He$^{+}$/H$^{+}$ show values higher than the average at the same region.  
It is worth noticing that, even though He$^{+}$/H$^{+}$ is a low-ionization
specie, a great amount of He$^0$ is also expected to be present in
this low ionization nebula. 
In spite of the peculiar behaviour of the 
low-ionization ions, as well as O$^{++}$/H$^{+}$ and He$^{+}$/H$^{+}$, the 
electron temperature maps (particularly relevant for abundance 
determinations) have no obvious gradient (see  corresponding map in
Figure~\ref{temden}) that would justify this result. 

Is this higher ionization tunnel of emission in NGC~40 similar to the jet 
structures of a number of other nebulae? In NGC~7009, for instance, two pairs 
of low-ionization knots are clearly seen, in addition to a pair of jets,
in which the excitation degree is significantly higher than that of the
immediate vicinity \citep{goncalves03}. However, contrary to what we
find for NGC~40, in the case of NGC~7009 the electron density of the
higher-ionization emission tunnel is a factor of 2 lower than in the
other nebular regions. But note that in both cases the \te\ is roughly 
the same throughout the nebula.

Concerning the element abundances of helium in NGC~40, He/H, we found the mean 
value of 7.08~$\times$~10$^{-2}$.  This result has been obtained by summing the 
He$^+$ and He$^{++}$ ionic abundance maps, and taking its mean value. 
This extremely low total helium abundance can be explained 
by the fact that a significant amount of He$^0$ is expected to be 
present in a low excitation nebula like NGC~40, and He$^0$ was not considered 
in the He/H given above.  
To circumvent this kind of problem, \citet{zhang03} suggested that an ICF 
based on the S$^+$ and S$^{++}$ abundances can be used to account for the 
He$^0$ abundance of low excitation PNe. Using our own data we were 
not able to calculate a S$^{++}$ abundance map, nevertheless, if we consider 
the S$^{++}$ abundances found by \citet{b17} and \citet{b9} 
(Table~\ref{results}), and the mean value of the S$^+$/H$^{+}$ map that 
we calculated (8.88~$\times$~10$^{-7}$, in Table~\ref{results}), we can 
determine a more reliable value for the He total abundance of NGC~40. Doing 
this exercise, we find He/H$_{(ICF - P)}$=9.32~$\times$~10$^{-2}$, adopting 
S$^{++}$/H$^{+}$ from Pottasch et al. (2003), and 
He/H$_{(ICF - L)}$=1.18~$\times$~10$^{-1}$ by using S$^{++}$/H$^{+}$ from Liu et 
al. (2004b). These three He/H are shown in Table~\ref{results}. Notice that the 
interval of S$^{++}$ abundances from \citet{b17} and \citet{b9} contains 
the mean S$^{++}$ abundance we found from our slit G, if we consider the NS, 
NIR, SIR and SS regions (again, that is not the case for the WN region, 
because of the contamination from the central star region). And, finally, not 
only we believe that 9.32~$\times$~10$^{-2}$ and 1.18~$\times$~10$^{-1}$ should 
correspond to the lower and upper He abundance limits of NGC~40, but they are 
also similar to the previous results found by \citet{b9}, who adopted the same 
ICF scheme to account for the unobserved amount of He$^0$ in the total helium 
abundance of NGC~40.

\begin{figure}
\begin{center}
\includegraphics[width = 0.48 \textwidth]{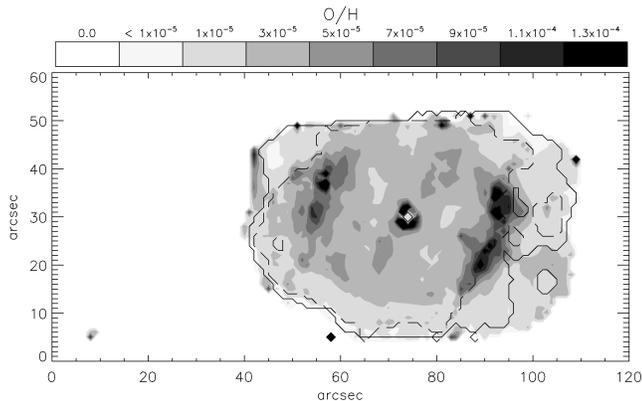}
\caption[]{Inferior limit of the oxygen chemical abundance map.}
\label{abund}
\end{center}
\end{figure}

The total oxygen abundance map is shown in Figure~\ref{abund}. This map
has been obtained summing the ionic abundance maps of O$^{0}$ and
O$^{++}$ --since, unfortunately, O$^{+}$ map is not available. This
element abundance is, in fact, only a lower limit. The  corresponding
mean value of the O/H map is 4.10~$\times$~10$^{-5}$; a much lower 
value when compared with the results from the literature. That is a 
consequence of the absence of the O$^{+}$ abundance map. As shown by 
\citet{b17} and \citet{b9}, this ion dominates the oxygen abundance of 
this nebula.

\bigskip

\section{Summary and Conclusions}

In this paper we presented a spatially-resolved analysis of the physical 
and chemical characteristics of the planetary nebula NGC~40. To reach 
these results, 16 long-slit parallel optical spectra were obtained, from 
which we created 31 emission-line spectroscopic maps.  These maps 
provide the spatially-resolved characteristics of the line 
emission, and the total fluxes for the whole nebula.

Using the usual diagnostic ratios, these maps were applied to calculate
the 2D density, temperature and abundance information for this
nebula. To derive these quantities, we adapted the tasks from the 
well-established {\sc iraf} \emph{nebular} package \citep{b5}, to work with
two-dimensional data, thereby developing, and proving the efficiency
of, the {\sc 2d\_neb} package.

Each emission-line map was corrected for reddening, pixel by pixel,
using the c(H$\beta$) map that was created from the H$\alpha$/H$\beta$, 
H$\gamma$/H$\beta$ and H$\delta$/H$\beta$ map ratios. The c(H$\beta$) map 
shows interesting features, with values that vary from less than 
$\sim$0.2 up to more than $\sim$0.6, as shown in the histogram of 
Figure~\ref{cbeta}, with a mean characteristic  of the whole nebula, 
c(H$\beta$) equal to 0.42. The above spatial extinction variation 
could indicate that dust is mixed with the ionized gas of NGC~40.

Our \tenii ~map shows only a slight variation from region to region, from
$\sim$8,000~K up to $\sim$9,500~K. On the other hand, these variations
are much less prominent than those found in the \nesii ~map, where the
values varies between $\sim$1,000~cm$^{-3}$ and
$\sim$3,000~cm$^{-3}$. The histograms of both physical parameters,
\tenii ~and \nesii, show this dispersion. The density map also
shows that there is a trend to lower density regions on the west side
of the nebula, but it does not show any trend for the outer regions 
compared to the inner ones. The mean \nesii ~(1,650~cm$^{-3}$) and
\tenii ~(8,850~K), from the spectroscopic maps, are in good agreement
with other results from the literature \citep{b6,b8,b17,b4,b2}.  On
the other hand, the maps show spatial variations that were not
addressed by previous works. 

Several of the ionic abundance maps show significant spatial variations.  i) The
Ar$^{3+}$/H$^+$ and Cl$^{++}$/H$^+$ maps show a ring structure on the inner 
region of NGC~40, and also high concentrations from these species on the outer 
region of the nebula. ii) The O$^0$/H$^+$ e O$^{++}$/H$^+$ maps show strong
spatial variations. In the first case, the contrast of the 
regions with O$^0$/H$^+$ $<$ 10$^{-5}$ is clear. For the second, most of the 
nebula has O$^{++}$/H$^+$ between 10$^{-5}$ and 3$\times$10$^{-5}$. 
However, regions with lower O$^{++}$ abundances are present in the outer 
regions of the nebula, and regions in which O$^{++}$/H$^+$ varies from 
$\sim$5$\times$10$^{-5}$ up to $\sim$1.3$\times$10$^{-4}$ are found, 
primarily between the inner rim and outer shell. iii) The He$^+$/H$^+$
abundance is also inhomogeneous. iv) The N$^{+}$/H$^+$ abundance map 
shows that almost the whole nebula has an N$^{+}$/H$^+$ between 
5$\times$10$^{-5}$ and 9$\times$10$^{-5}$ or that N$^{+}$/H$^+$ does not 
vary significantly throughout the nebula. v) Finally, the S$^{+}$/H map 
shows clumps, in which the S$^{+}$/H varies between less than 4$\times$10$^{-7}$ 
up to more than 10$^{-6}$.

It is important to point out that maps represent a projection of 
the nebula along the line of sight on the sky, and so nothing can be said 
about the tridimensional state of the gas (T$_e$, N$_e$, ionic as well as 
total abundances) from our analysis of NGC40.

However, the results obtained in this work allow us to conclude that the 
spatial variations found in the c(H$\beta$), \tenii, \nesii, and 
abundance maps unequivocally confirm the necessity of an analysis with 
spatial resolution for a more complete study of the physical and chemical
properties of planetary nebulae. Furthermore, the spectroscopic
mapping technique used in this paper and in previous works shows
important advantages when compared with the usual technique that uses
one or a few long-slits to analyse the object. As an extra ingredient 
in this discussion, we note that \citet{corradi97},
\citet{perinotto98}, and \citet{goncalves03,goncalves06} did not find
significant spatial variation of the chemical abundances when
analysing a number of other PNe. As we do find these variations in
NGC~40, an important question that remains to be addressed is: if
analysed with the spectroscopic mapping technique, as in this work, would 
those $\sim$20 PNe studied with single long-slit spatially resolved spectra
(above references) show significant chemical variations?  

  Also in the context of distant extragalactic PNe, it is clear
  that spatially resolved data using techniques similar to the one
  applied here to study NGC~40 are desirable. However, our
  telescopes/instruments prevent us of such a detailed study of
  distant objects. Fortunately in these situations most of the
  emission of a PN is captured in a single slit observation --making
  the average values more representative. And, therefore, the
  interpretation of results for distant extragalactic PNe are, at
  least in a statistical sense, not as seriously affected by the lack of
  spatial resolution.

\section*{Acknowledgments}

We would like to dedicate this paper to the memory of Hugo E. Schwarz, the 
observer of the data used here. We also thank John Danziger, the referee, for 
his revision of the paper with suggestions that helped to improved it. Three 
Brazilian agencies gave us partial support for this work. So MLLF, DRG and HM 
would like to thank CAPES, FAPERJ's E-26/110.107/2008 grant and FAPESP 
(2003/09692-0 grant), respectively.

\bsp

\label{lastpage}

\end{document}